\documentclass[10pt,letterpaper]{article}
\usepackage[top=0.85in,left=2.75in,footskip=0.75in]{geometry}

\usepackage{amsmath,amssymb}

\usepackage{changepage}

\usepackage{textcomp,marvosym}

\usepackage{cite}

\usepackage{nameref}
\usepackage[hidelinks]{hyperref}

\usepackage[right]{lineno}

\usepackage[nopatch=eqnum]{microtype}
\DisableLigatures[f]{encoding = *, family = * }

\usepackage[table]{xcolor}
\usepackage{makecell}

\usepackage{array}

\newcolumntype{+}{!{\vrule width 2pt}}

\newlength\savedwidth

\raggedright
\usepackage[aboveskip=1pt,labelfont=bf,labelsep=period,justification=raggedright,singlelinecheck=off]{caption}
\renewcommand{\figurename}{Fig}

\makeatletter
\renewcommand{\@biblabel}[1]{\quad#1.}
\makeatother

\usepackage{lastpage,fancyhdr,graphicx}
\usepackage{epstopdf}
\fancyheadoffset[L]{2.25in}
\fancyfootoffset[L]{2.25in}
\begin{document}
\vspace*{0.2in}

\begin{flushleft}
{\Large
\textbf\newline{Optimizing chemoradiotherapy for malignant gliomas: a validated mathematical approach} 
}
\newline
\\
Miguel Perales-Pat\'on \textsuperscript{1,3,4\Yinyang},
Matteo Italia \textsuperscript{1,3,4\Yinyang*},
Mar\'ia Castell\'o-Pons \textsuperscript{2,5\Yinyang},
Irene G\'omez-Soria \textsuperscript{2},
Juan Belmonte-Beitia\textsuperscript{1,3,4},
Pilar S\'anchez-G\'omez\textsuperscript{2},
V\'ictor M. P\'erez-Garc\'ia\textsuperscript{1,3,4}
\\
\bigskip
\textbf{1} Mathematical Oncology Laboratory (MOLAB), Instituto de Matem\'atica Aplicada a la Ciencia y la Ingenier\'ia, University of Castilla-La Mancha, Spain
\\
\textbf{2} Neurooncology Unit, Unidad Funcional de Investigaci\'on de Enfermedades Cr\'onicas (UFIEC), Instituto de Salud Carlos III (ISCIII), Madrid, Spain
\\
\textbf{3} Department of Mathematics, Escuela T\'ecnica Superior de Ingenier\'ia Industrial, University of Castilla-La Mancha, Spain
\\
\textbf{4} Laboratorio de Oncolog\'ia Matem\'atica, Instituto de Investigaci\'on Sanitaria
de Castilla-La Mancha (IDISCAM), Spain
\\
\textbf{5} PhD Programme on Biomedical Sciences and Public Health, Universidad Nacional de Educaci\'on a Distancia, UNED-ISCIII, Madrid, Spain
\\
\bigskip

%
%
\Yinyang These authors contributed equally to this work.





* matteo.italia@uclm.es

\end{flushleft}
\section*{Abstract}
Malignant gliomas (MGs), particularly glioblastoma, are among the most aggressive brain tumors, with limited treatment options and a poor prognosis. Maximal safe resection and the so-called Stupp protocol are the standard first-line therapies. Despite combining radiotherapy and chemotherapy in an intensive manner, it provides limited survival benefits over radiation therapy alone, underscoring the need for innovative therapeutic strategies. Emerging evidence suggests that alternative dosing schedules, such as less aggressive regimens with extended intervals between consecutive treatment applications, may improve outcomes, enhancing survival, delaying the emergence of resistance, and minimizing side effects. In this study, we develop, calibrate, and validate in animal models a novel ordinary differential equation-based mathematical model, using \textit{in vivo} data to describe MG dynamics under combined chemoradiotherapy. The proposed model incorporates key biological processes, including cancer cell dormancy, phenotypic switching, drug resistance through persister cells, and treatment-induced effects. Through \textit{in silico} trials, we identified optimized combination treatment protocols that may outperform the standard Stupp protocol. Finally, we computationally extrapolated the results obtained from the \textit{in vivo} animal model to humans, showing up to a four-fold increase in median survival with protracted administration protocols \textit{in silico}. Although further experimental and clinical validation is required, our framework provides a computational foundation to optimize and personalize treatment strategies for MG and potentially other cancers with similar biological mechanisms.

\section*{Author summary}
Malignant gliomas are highly aggressive brain tumors with limited treatment options and a poor prognosis. Standard treatment consists of surgical resection followed by chemoradiotherapy; however, its efficacy is often hindered by tumor resistance.
Our model simulates tumor growth and treatment responses to radiotherapy, chemotherapy, and their combination. A key contribution of our work is the explicit integration of radiotherapy effects, which enhances our understanding of treatment interactions. The model was calibrated and validated using experimental data from mice, allowing us to test different treatment scenarios in virtual trials.
Our simulations identified an alternative treatment protocol that improved survival outcomes in mice compared to the standard protocol. By scaling the model to human patients, \textit{in silico} trilas show that optimal treatments can have a significant impact on survival, potentially reducing treatment-related side effects and improving quality of life.
These findings underscore the potential of mathematical modeling as a powerful tool for optimizing cancer treatments. By offering a framework for evaluating protocols prior to clinical trials, our work contributes to the advancement of more effective and personalized approaches to the treatment of malignant gliomas.


\section*{Introduction}\label{sec:introduction}
Malignant gliomas (MGs) are aggressive tumors that resemble glial cells, the supportive cells of the central nervous system. Despite significant advances in the treatment of many cancer types, progress in MG treatment has remained limited \cite{santucci2020progress}. Glioblastoma (GBM), the most prevalent and aggressive form of MG, is a malignant primary brain tumor associated with a poor prognosis and few effective treatment options. The current standard of care, known as the Stupp protocol, combines maximal surgical resection with chemoradiation therapy, using temozolomide (TMZ) as chemotherapy (CT) \cite{stupp2005radiotherapy}.

TMZ is an alkylating agent that crosses the blood-brain barrier and has a favorable toxicity profile \cite{agarwala2000temozolomide}.
It exerts its cytotoxic effects primarily by inducing DNA methylation, resulting in the formation of O6-methylguanine lesions. These lesions induce DNA mismatches during replication, leading to double strand breaks and cell death, particularly in actively proliferating tumor cells \cite{roos2007apoptosis}. The so-called Stupp protocol consists of maximal surgical resection followed by fractionated radiation therapy (RT) with concurrent daily TMZ for six weeks, and subsequent monthly adjuvant TMZ cycles \cite{stupp2005radiotherapy}. This  extends overall survival (OS) by only about three months compared to RT alone, highlighting the need for more effective treatment strategies. Given the short half-life of TMZ of approximately two hours in humans \cite{rudek2004temozolomide}, increasing the dose or frequency of administration has been proposed as strategies to enhance the efficacy of this CT by narrowing the DNA repair window \cite{wick2009new}. However, such approaches may also drive the emergence of resistance mechanisms, potentially undermining the drug's therapeutic efficacy \cite{singh2021mechanisms}. Optimizing TMZ administration is therefore crucial to maximize cytotoxicity while minimizing resistance development.

Emerging evidence suggests that aggressive treatment regimens, like the Stupp protocol, may not be universally effective across all cancer types or patient populations \cite{sachs2016optimal, italia2023mathematical}. In the context of GBM, slower-growing tumors or those with lower proliferative indices may benefit from alternative treatment approaches. 
Recent studies indicate that regimens with extended intervals between TMZ doses may outperform standard protocols in mouse models of these tumors \cite{arias2017metabolomics,ferrer2017metronomic,calero2021immune,wu2020anti}, particularly in mice with slow-growing GBMs \cite{segura2022optimal}. These protracted schedules enhance OS by aligning treatment delivery more closely with the biological characteristics of the tumor, thereby mitigating the development of TMZ resistance. Other alternative treatment strategies, such as the metronomic schedule -- characterized by the continuous or regular administration of low-dose chemotherapy over extended periods -- have also gained attention as promising alternatives to the standard Stupp protocol. This approach is supported by both mathematical modeling studies \cite{pasquier2010metronomic, faivre2013mathematical, benzekry2015metronomic} and retrospective clinical trials \cite{cominelli2015egfr}.

In oncology research, the rapid testing and refinement of therapeutic strategies are crucial for advancing patient care \cite{mercieca2018importance}. However, experimental studies and clinical trials are typically time-consuming and resource-intensive, which can hinder innovation. Mechanistic modeling, particularly through ordinary differential equations (ODEs), offers a computationally efficient alternative \cite{alfonso2020translational}. ODE-based mechanistic models are computationally efficient, enabling the simulation of multiple treatment scenarios in a fraction of the time required by other methods, such as agent-based models. These models are particularly valuable for studying tumor dynamics and treatment evolution by providing large-scale virtual cohorts that can mimic the variability observed in real-world populations \cite{dormand2018numerical}. \textit{In silico} trials, or virtual simulations of clinical trials, have emerged as a powerful tool in mathematical oncology \cite{gevertz2024assessing}, helping de-risk clinical trial outcomes and guide the design of focused, high-likelihood experimental studies \cite{brown2022derisking}. By simulating various treatment regimens, researchers can systematically assess their effects and identify the most effective strategies for specific patient subgroups, thus accelerating discovery while reducing costs and ethical concerns \cite{brown2022derisking}.

There exists a wide range of MG mathematical models, including ODEs, partial differential equations (PDEs), and cellular automata (CA), each capturing distinct facets of tumor biology. ODE models have been developed to investigate features such as immunotherapy \cite{khajanchi2021impact}, tumor dynamics prior to diagnosis \cite{sturrock2015mathematical}, and tumor progression using image-derived data \cite{khajanchi2021impact} and volumetric longitudinal data \cite{bogdanska2017mathematical}. PDE-based models also exhibit considerable diversity. As some examples, Swanson et al. investigated angiogenesis \cite{swanson2011quantifying}, Gerlee et al. modeled tumor growth \cite{gerlee2016travelling}, Stein et al. optimized treatment regimens involving lapatinib \cite{stein2018mathematical}, Corwin et al. explored patient-specific radiotherapy regimens \cite{corwin2013toward}, Swan et al. modeled tumor spread \cite{swan2018patient}, Neal et al. studied survival outcomes \cite{neal2013discriminating}, Jackson et al. informed clinical practice \cite{jackson2015patient}, and Gerlee et al. looked at phenotypic switching in growth and invasion \cite{gerlee2012impact}. Similarly, CA models have been instrumental in exploring biological phenomena such as cell migration and invasion \cite{aubert2006cellular, hatzikirou2012go, tektonidis2011identification}, as well as intra-tumoral heterogeneity in GBM \cite{saucedo2024simple, jimenez2021mesoscopic}. 

In conclusion, mathematical models provide a valuable framework for understanding and optimizing cancer treatment strategies by abstracting and formalizing biological systems. However, many existing models focus on either limited biological processes, short-term predictions, or do not account for the complexity of resistance mechanisms \cite{ayala2021optimal,perez2019computational}. Resistance mechanisms, such as the emergence of drug-tolerant persister populations \cite{segura2022optimal} and phenotypic transitions \cite{delobel2023overcoming}, are critical for understanding treatment failure in MGs and other cancers.
Furthermore,  models often lack experimental calibration and validation \cite{sujitha2023fractional}, or focus on a single type of therapy, limiting their applicability to combination regimens \cite{segura2022optimal}. Some models proposed schedules that are challenging to implement clinically, such as the one suggested by Leder et al. \cite{leder2014mathematical}, which emphasized radiotherapy but overlooked critical factors like accelerated tumor repopulation \cite{accelrepopafterRT1, accelrepopafterRT2, accelrepopafterRT3} and suggested schedules that involve non-standard irradiation patterns or irregular treatment regimens that would affect the planning of radiation oncology services.

In this study, we developed, calibrated, and validated an ODE-based compartmental model integrating key biological processes involved in the response of MGs to combined TMZ and RT treatment. This model incorporates the biology governing the emergence of TMZ resistance through persister cells \cite{vallette2019dormant,rabe2020identification}, while also describing phenotypic transitions \cite{dormantstateandback}, actively proliferating tumor cells \cite{dahlrot2021prognostic}, and the effects of treatments on tumor dynamics \cite{lomax2013biological, agarwala2000temozolomide}.
We used this validated model to virtually explore optimal treatment strategies and translate these findings into human treatment protocols. Our results demonstrate that certain strategies can outperform the Stupp protocol in \textit{in silico} trials, leading to up to a fourfold increase in survival, although further experimental and clinical validation is required.

Note that while the primary focus is on MGs, the framework can be adapted and calibrated for other cancers exhibiting similar biological mechanisms, such as resistance development through persistence under CT. In fact, persister subpopulations, first observed in the resistance of microorganisms to antibiotics \cite{bacteriapersisters}, have been identified in different cancers \cite{ramirez2016diverse}, including lung \cite{paramgeneracionpersisters, oren2021cycling}, melanoma \cite{rambow2018toward,persisters1}, breast \cite{hangauer2017drug}, and leukemia \cite{boyd2018identification,ho2016evolution}.

\section*{Materials and methods}\label{sec:model}


\subsection*{Ethical statement}\label{subsec:ethicalstatement}
Animal experiments were reviewed and approved by the Research Ethics and Animal Welfare Committee at ``Instituto de Salud Carlos III" (\texttt{PROEX 306.7/22}), in agreement with the European Union and national directives. Intracranial transplantation ($250000$ cells) was done as previously described in \cite{segura2021tumor, gargini2020idh}.

\subsection*{Biological background}\label{subsec:biologicalhypothesis}

We build here an ordinary differential equations (ODE) model of MG growth and response to combined radiotherapy (RT) and chemotherapy/temozolomide (CT/TMZ) treatment. This model relies on a compartmental description of the temporal evolution of several genetically/phenotypically homogeneous tumor cell subpopulations and their response to RT and CT.  Specifically, the model includes sensitive ($S$), quiescent ($Q$), damaged ($D$), persister ($P$), and resistant ($R$) cancer cell subpopulations. For simplicity, the model does not consider either spatial dependencies or stochastic effects.

We assume that sensitive cells ($S$) exhibit a proliferative phenotype, grow at a characteristic rate, and are sensitive to both CT \cite{chemodamage} and RT \cite{radiodamage}. Quiescent cells ($Q$) are characterized by a non-proliferative phenotype; therefore, we assume that they cannot be damaged either by CT or RT. It is well established that quiescent cancer cells can be stimulated to proliferate by RT, leading to the so-called accelerated repopulation after sublethal damage in cancer cells treated with RT \cite{accelrepopafterRT2,accelrepopafterRT1,accelrepopafterRT3}. We also assume that cancer cells spontaneously change their phenotypes from proliferative to quiescent and vice-versa \cite{dormantstateandback}, resulting in an equilibrium between both populations governed by the rates of phenotype changes \cite{ayala2021optimal}.

The damaged cell subpopulation ($D$) consists of viable cancer cells whose DNA has been significantly affected or lethally damaged by treatments. When these cells attempt to divide, they undergo mitotic catastrophe and subsequently die \cite{ hirose2005akt, vakifahmetoglu2008death, programmedcelldeathglioma}. This late death mechanism following low-dose irradiation is the dominant one within the dose ranges used in fractionated RT \cite{joiner2025basic}.

The persister population is characterized by a dormant, non-proliferative state. This phenotype is induced in proliferative, drug-sensitive cells in response to CT exposure. We assume that the transition of sensitive cells to a resistant state occurs in two steps, simplifying a more complex and continuous process involving the increase in levels of resistance markers \cite{rabe2020identification}. Under drug exposure, sensitive cells initially enter an intermediate and transient persister state ($P_I$) with a finite lifetime. Further exposure to the drug stabilizes these persister cells into a fully persister phenotype ($P$). Continued exposure to additional chemotherapy doses induces their transition to a fully resistant, permanent phenotype ($R$). In the absence of drug exposure, both phenotypes of persister cells revert to a sensitive phenotype after a characteristic time \cite{adamski2017dormant, vallette2019dormant}.

The resistant cell population ($R$) is characterized by minimal drug sensitivity, which we assumed to be negligible in this model \cite{rabe2020identification}. However, due to their proliferative nature, they are assumed to be susceptible to RT \cite{lomax2013biological}.  

Finally, for simplicity, we assume that RT damages instantaneously proliferative cancer cells \cite{lomax2013biological}, since irradiation times are typically much shorter than all of the other cellular times included in the model. It also induces quiescent cells to turn proliferative in very short periods of time, as will be described mathematically later. Conversely, CT targets actively proliferating sensitive cancer cells, with its effect depending on the actual drug concentration, which follows first-order pharmacokinetics \cite{agarwala2000temozolomide}.

\subsection*{Model equations}\label{subsec:themodelequations}
Fig \ref{fig:model} presents a graphical scheme illustrating the key biological variables included in the model and their interactions. The ODE system that governs the dynamics is given by:
\begin{figure}[ht]
\centering
\includegraphics[width=0.9\textwidth]{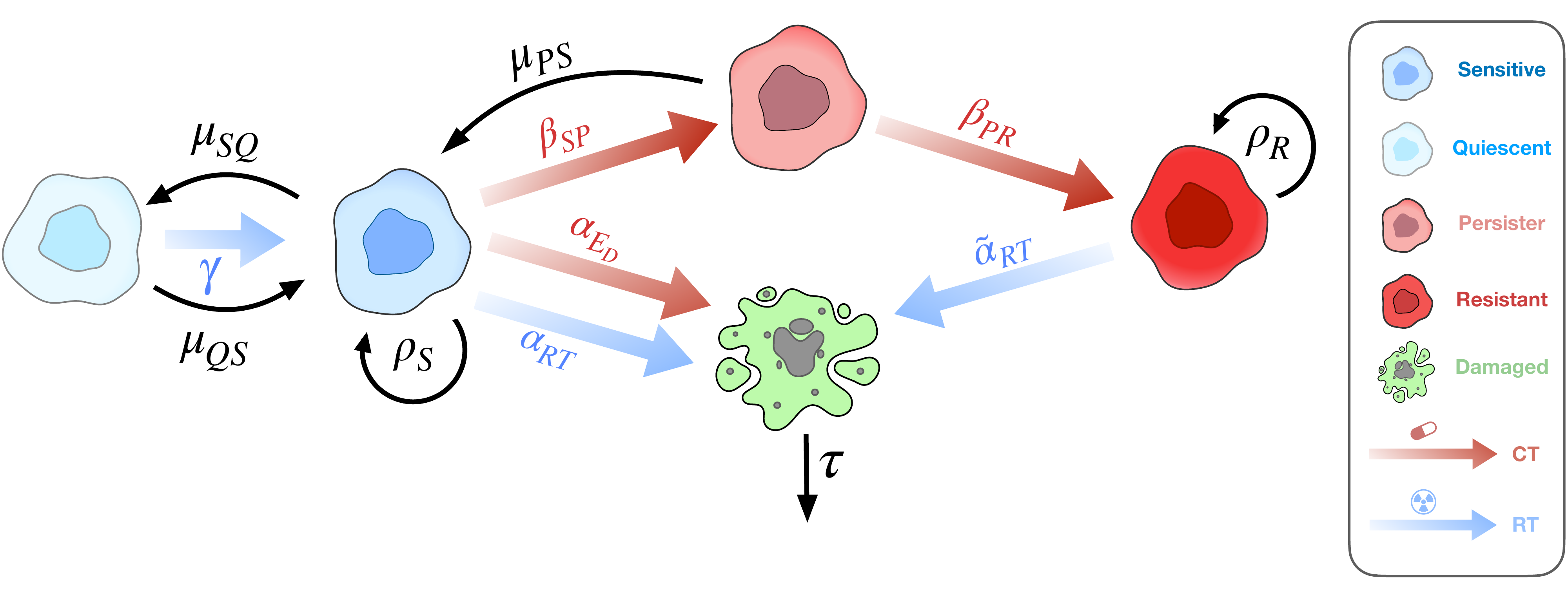}
\caption{A graphical representation of the model highlighting the tumor subpopulations, their dynamics, and the effects of RT and TMZ.}\label{fig:model}
\end{figure}

\begin{subequations} \label{eq:systemEDO}
    \begin{align}
        \dot{S} &= \rho_S S- \mu_{SQ} S + \mu_{QS} Q - \beta_{SP_I}SE_D  +\mu_{PS} P - \alpha_{E_D}SE_D,  \label{eq:S} \\
        \dot{Q} &= \mu_{SQ} S - \mu_{QS} Q ,  \label{eq:Q} \\
        \dot{P}_I &= \beta_{SP_I}ST - f(E_D)P_I,  \label{eq:PI} \\
        \dot{P} &= f(E_D)P_I - \mu_{PS} P - \beta_{PR} PE_D, \label{eq:P} \\
        \dot{R} &= \rho_R R + \beta_{PR} PE_D,  \label{eq:R} \\
        \dot{D} &= \alpha_{E_D}SE_D-\tau D,\label{eq:D}        \\
        \dot{E}_D &=-\lambda E_D,                                           \label{eq:TMZ}
    \end{align}
\end{subequations}

where \begin{equation}\label{eq:activationfunction}
    f(E_D)=7.5\left(1-\tanh{\frac{E_D-0.01}{0.01}}\right).
\end{equation}

Eq \eqref{eq:S} describes the dynamics of sensitive cells ($S$). In this equation, $\rho_S$ represents the population proliferation rate, $\mu_{SQ}$ governs the transition from this compartment to the quiescent cell compartment ($Q$), $\mu_{QS}$ the reverse transition from $Q$ to $S$, $\beta_{SP_I}$ denotes the induction of persisters due to the drug, $\mu_{PS}$ represents the reverse flow from persister cells ($P$) to sensitive cells ($S$) that is independent of the drug, and $\alpha_{E_D}$ indicates the DNA damage caused by the drug.

Eq \eqref{eq:Q} models the dynamics of quiescent cells ($Q$), where $\mu_{SQ}$ and $\mu_{QS}$ govern the flows from and to the sensitive cell compartment ($S$), respectively.

Eq \eqref{eq:PI} describes the dynamics of the intermediate and transient persister state ($P_I$), induced in sensitive cells upon first exposure to the drug at a rate $\beta_{SP_I}$. Subsequently, the activation function $f(E_D)$ stabilizes the transient ($P_I$) into a fully persister phenotype ($P$). Note that this function $f(E_D)$ should be designed to mathematically capture the previously described biological behavior: newly formed (intermediate and temporary) persister cells cannot immediately transition to resistant cells within the same dose application. To reflect this dynamic, the activation function should remain close to zero for high values of $E_D$ and become positive for low values of $E_D$, thereby allowing the transition to the fully persister phenotype only under the latter condition. A classical sigmoid function, such as the one in Eq \eqref{eq:activationfunction}, was adopted following Delobel et al. work \cite{delobel2023overcoming}. 

Eq \eqref{eq:P} governs the dynamics of the fully persister state ($P$). These cells revert to a sensitive phenotype at a rate $\mu_{PS}$ or transition to a resistant state ($R$) at a rate $\beta_{PR}$ depending on the drug efficacy ($E_D$).

Eq \eqref{eq:R} models the dynamics of resistant cells ($R$), where $\rho_R$ represents their constant proliferation rate. The transition from the fully persister compartment ($P$) to the resistant compartment ($R$) is governed by $\beta_{PR}$, which depends on the current drug efficacy ($E_D$).

Eq \eqref{eq:D} describes the dynamics of the damaged population ($D$), where $\alpha_{E_D}$, which depends on the actual drug efficacy ($E_D$), governs the entering flow from sensitive cells ($S$) due to drug damage, and $\tau$ represents the rate at which damaged cells die and are eliminated from the tumor mass.

Finally, Eq \eqref{eq:TMZ} describes the exponential decay of the drug ($E_D$) following first-order kinetics.

We denote the initial tumor size as $T_0:=T(t=0)$, which is given by the sum of all tumor subpopulations: $T_0=S(0)+Q(0)+P_I(0)+P(0)+R(0)+D(0)$. Initially, the environment is typically initially drug-free, $E_D(0)=0$. The total initial tumor size, its distribution among different tumor subpopulations, and the initial drug efficacy ($E_D(0)$) depend on the simulated context. For instance, before treatment starts, damaged, persister, and resistant cells are absent. In such scenarios, the initial conditions for Eqs \eqref{eq:systemEDO} are given by:
\begin{subequations} \label{Ki670}
\begin{eqnarray}
 P_I(0) & = & P(0)=R(0)=D(0)=E_D(0)=0, \\
 S(0) & = & \text{Ki-}67\cdot T_0, \\
 Q(0) & = & (1-\text{Ki-}67)\cdot T_0.
 \end{eqnarray}
 \end{subequations}
The value of $T_0$ can be inferred from volumetric imaging measurements in practical applications.
Simulations are terminated at predefined final times, which depend on the biological context, such as a specified mechanism or experimental duration, or significant events, such as the tumor reaching a critical size threshold (representing fatal tumor volume) or complete remission.

The model was solved and fitted to the available longitudinal volumetric data using \texttt{ode45} and \texttt{fmincon} functions, respectively, included in the scientific software package \texttt{MATLAB} (R$2023$b, The MathWorks, Inc., Natick, MA, USA), running on macOS Ventura $13.6.9$, with an Apple M2 Ultra chip.

\subsection*{Modeling treatment effects}

Treatments are assumed to be delivered instantaneously, as their typical time durations are much shorter than the longitudinal changes observed in tumor dynamics. For RT, the course is administered in around 10 minutes, while TMZ reaches peak concentrations in the brain in about 30 minutes. As in previous studies \cite{delobel2023overcoming,italia2023mathematical,otero2022dynamics,otero2024dynamics}, we temporarily stopped time integration of Eqs \eqref{eq:systemEDO} at treatment times and updated the variables involved in the treatment. Subsequently, time integration of Eqs \eqref{eq:systemEDO} is resumed with the updated state variables.
More specifically, RT induces DNA damage in proliferative cancer cells and stimulates quiescent cells to re-enter an actively proliferative state. Consequently, the evolution of subpopulations is instantaneously modified at treatment times $t_{i}$ of each irradiation dose $i$, with $i=1,2,\dots, N_{RT}$, where $N_{RT}$ denotes the total number of RT individual doses. Let $t_i^+:=\lim_{t\to t_i^+} t,\text{ and } t_i^-:=\lim_{t\to t_i^-} t,$ then the RT effects are modeled as follows:

\begin{subequations} \label{eq:systemRT}
    \begin{align}
S(t_i^+)&=(1-\alpha_{RT})S(t_i^-) + \gamma Q(t_i^-),\label{saltoS}\\
Q(t_i^+)&=(1-\gamma)Q(t_i^-),\label{saltoQ}\\
R(t_i^+)&=(1-\tilde{\alpha}_{RT})R(t_i^-),\label{saltoR}\\
D(t_i^+)&=D(t_i^-)+\alpha_{RT}S(t_i^-)+\tilde{\alpha}_{RT}R(t_i^-),\label{saltoD}
    \end{align}
\end{subequations}
where $\alpha_{RT}$ and $\tilde\alpha_{RT}$ represent the fractions of sensitive and drug-resistant populations, respectively, that are damaged by RT, and $\gamma$ denotes the fraction of quiescent cells stimulated into a proliferative state.

We will assume here that the administration of CT immediately affects the state of the drug concentration ($E_D$). TMZ reaches its highest concentrations within $30$ to $90$ minutes in humans \cite{neuropharmacokinetics} and in less than $1$ hour in mice \cite{ballesta2014multiscale}, a timescale significantly shorter than the tumor growth dynamics. Consequently, for the sake of simplicity, we assumed tumor sizes to remain constant between the time of TMZ administration and the time at which its peak concentration is reached. Thus, the time required to achieve maximum TMZ efficacy is considered instantaneous and coincides with the time of TMZ administration. The evolution of drug efficacy at time $\tilde t_j$ of dose administration $j$, with $j=1,2,\dots,N_{CT}$, where $N_{CT}$ represents the total number of CT doses, is modeled as follows:

\begin{equation} \label{saltoT}
E_D(\tilde t_j^+)=E_D(\tilde t_j^-)+\delta_{E_{D_j}}, 
\end{equation}

where $\delta_{E_{D_j}}$ represents the normalized administered dose $j$ ($0<\delta_{E_{D_j}}\leq1$), as modeled in \cite{delobel2023overcoming}.

\subsection*{Model parameters}\label{subsec:modelparameters}

Model parameters are typically derived from biological distributions, or calibrated on experimental data. In any case, when biological information from literature or experiments is available, it should be used to constrain parameter values and their relationships. 

MGs are known to exhibit varying levels of Ki-67, which represent the tumor's proliferative activity, specifically the percentage of cells that are currently actively proliferating \cite{dahlrot2021prognostic}. Although Ki-67 can be temporarily altered by treatments, it is found to be almost constant in the case of MGs and their basal values are recovered after treatments \cite{akkari2020dynamic}. To maintain a constant Ki-67, the fraction of proliferating cells should be related to the values of the transition rates between proliferating and quiescent populations as follows
 \begin{equation}\label{eq:flowQS}
\mu_{QS}=\text{Ki-}67\left(\frac{\mu_{SQ}}{1-\text{Ki-}67} - \rho_S\right),
\end{equation} as described by Ayala et al. \cite{ayala2021optimal}. As a consequence, to maintain non-negativity for all parameters, the constraint $\mu_{SQ}\geq\rho_S(1-\text{Ki-}67)$:=$\mu_{SQ_\text{min}}$ must be satisfied.

The growth rates of both sensitive and drug-resistant cells are highly heterogeneous
\cite{rutter2017mathematical}. \textit{In vivo} experiments have shown that drug-resistant MGs may grow more slowly \cite{yuan2018abt}, faster \cite{yang2021enhanced}, or at a similar rate \cite{gupta2014discordant} than the corresponding sensitive MG \textit{in vivo} models. For the sake of simplicity, we assumed here the same growth rate for sensitive and drug-resistant MG cells.

The induction of drug resistance in MGs has been investigated in both mice \cite{paramgeneracionpersisters,segura2022optimal} and humans \cite{delobel2023overcoming}. However, it is expected that the timescales of phenotypic changes, characteristic of tumor cells, remain of the same order in both humans and mice. Nevertheless, tumor demographic processes differ between species and occur on distinct timescales. Therefore, the persister cells generation rate ($\beta_{SP_I}$) and the resistant cells generation rate ($\beta_{PR}$) can be constrained by the minimum and maximum values observed in Delobel et al. work \cite{delobel2023overcoming}. 

The rate at which fully persister cells transition into sensitive cells ($\mu_{PS}$) has been characterized in a previous study using the same \textit{in vivo} animal model \cite{segura2022optimal}.

Note that, since RT and TMZ induce cell death through mitotic catastrophe after completing one or several mitosis cycles, the time constant for the removal of damaged cancer cells from the system should be of the same order as the proliferative parameter, or smaller. This indicates that cancer cells attempt to divide at least once before dying \cite{joiner2025basic}. In a previous study on low-grade MGs, the average number of mitotic cycles required to induce apoptosis after RT was considered to be 3 \cite{perez2015delay}. In this study, to model a broader range of MGs, we set the elimination rate ($\tau$) interval with a lower bound of one-hundredth and an upper bound of twice the proliferative rate ($\rho_S$).

TMZ has been reported to have a half-life of $1.4$ hours in mice \cite{cho2020pharmacokinetic}, allowing us to estimate the exponential decay $\lambda$ of TMZ efficacy ($E_D$) as $\left(\lambda=\log(2)/1.4=0.4951\text{ h}^{-1}= 11.8825 \text{ day}^{-1}\right)$. 
 
The information regarding the \textit{in vivo} model parameters is summarized in Table \ref{tab:param}.

\begin{table}[!ht]
\centering 
\caption{Summary of the model parameters describing MG growth in mice, including symbol, range, unit, meaning, and sources. Note that the parameters in the top, middle, and bottom row groups represent demographic processes, response to radiotherapy (RT), and response to temozolomide (TMZ), respectively. Some parameter ranges are constrained to their feasible values (FV). Parameters are structures in three blocks, the first one being treatment independent, the second one related to the response to RT and the third one to the response to TMZ.}
\begin{tabular}{ccccc}
\textbf{Parameter} & \textbf{Meaning} & \textbf{Range} & \textbf{Unit} & \textbf{Source}\\ \hline & & & & \\
{$\rho_S$} & \makecell{Sensitive population \\ proliferation rate} & {$[0,3]$} & {d$^{-1}$} &  \cite{oraiopoulou2017vitro}\\ 
{$\rho_R$} & \makecell{Drug-resistant population \\ proliferation rate} & {$[0,3]$} & {d$^{-1}$} & \cite{gupta2014discordant} \\ 
$\mu_{SQ}$ & \makecell{Mutation rate from \\ sensitive to quiescent} & $\geq \mu_{SQ_{\text{min}}}$ & d$^{-1}$ & FV \\ 
$\mu_{QS}$ & \makecell{Mutation rate from \\ quiescent to sensitive} & $f(\mu_{SQ})$ & d$^{-1}$ & \cite{ayala2021optimal} \\ 
$\mu_{PS}$ & \makecell{Mutation rate from \\ persister to sensitive} & $0.2$ & d$^{-1}$ & \cite{segura2022optimal} \\
Ki-67 & \makecell{Fraction of \\ proliferating cells} & $0.1$ & - & \cite{segura2022optimal}\\  & & & & \\  \hline & & & & \\ 
$\gamma$ & \makecell{Proliferation induced in \\ quiescent population by RT} & $[0,1]$ & - & FV \\ 
$\alpha_{RT}$ & \makecell{Fraction of sensitive \\ population damaged by RT} & $[0,1]$ & - & FV \\ 
$\tilde\alpha_{RT}$ & \makecell{Fraction of drug-resistant \\ population damaged by RT} & $[0,1]$ & - & FV \\ 
$\tau$ & \makecell{Damaged population \\ elimination rate} & $\left[\frac{\rho_S}{100},2\rho_S\right]$ & d$^{-1}$ & \cite{joiner2025basic} \\ & & & & \\  \hline & & & & \\ 
$\beta_{SP_I}$ & \makecell{Persister cells \\ generation rate} & $48$ & d$^{-1}$ & \cite{paramgeneracionpersisters} \\ 
$\beta_{PR}$ & \makecell{Resistant cells \\ generation rate} & $16$ & d$^{-1}$ & \cite{paramgeneracionpersisters} \\ 
$\alpha_{E_D}$ & \makecell{TMZ killing efficiency} & $\geq0$ & d$^{-1}$ & FV \\ 
$\lambda$ & \makecell{TMZ efficacy \\ decay rate} & $11.8825$ & d$^{-1}$ & \cite{cho2020pharmacokinetic} \\ 
$\delta_{E_{D_j}}$ & \makecell{TMZ efficacy increment \\ of dose $j$} & $[0,1]$ & - & FV \\ & & & & \\  \hline 
\end{tabular}
\label{tab:param}
\end{table}

When animal- or patient-specific data are available, the model can be personalized to virtually represent the actual animal or patient, thereby creating what is known as a digital twin \cite{wu2022integrating}. This virtual representation enables the testing and prediction of real system behavior under various potential scenarios through simulations. The unique characteristics of each patient can be incorporated into our model via a set of parameters. MGs, like other brain tumors, exhibit significant inter-patient variability. Consequently, each patient is expected to be represented by a distinct set of individualized parameter values.

To personalize the model for individual patients, the optimal set of parameter values must be determined by minimizing the discrepancy between actual data and model predictions. Specifically, if the available data correspond to the total tumor size or an appropriate proxy, the error between the model-simulated tumor volume ($T(t_j) = S(t_j)+Q(t_j)+P_I(t_j)+P(t_j)+R(t_j)+D(t_j)$) and the observed longitudinal data ($T_j$) at time points $t_j$, with $j = 0,1, \dots,N$, should be minimized. These time points represent consecutive clinical observations. The model calibration process involves fitting all parameters to the total tumor evolution over time. The results of this calibration procedure are presented in Section \nameref{sec:RT_calibration}.

Note that it is possible to fit a specific subset of all parameters using \textit{ad hoc} experiments and/or in subsequent steps. 
For example, parameters related to proliferative and quiescent populations can be adjusted using data from experiments of tumor growth without treatments. Following this approach, the parameters related to RT or CT can be adjusted with data where only these treatments are applied. The fits can be implemented as numerical optimization processes. To ensure the robustness of the fit, several initial random seeds within the prescribed bounds should be investigated. This increases the likelihood that the result of the optimization process will be a global minimum (best fit), rather than a local minimum. 

\subsection*{Experimental calibration and validation}\label{subsec:experimentalmethods}
 \nameref{subsec:themodelequations}, several \textit{in vivo} experiments were conducted using the mouse subventricular zone (SVZ) mouse model, developed through retroviral expression of EGFR-wt in primary cultures of neural stem cells from mice. The characterization of the model and the protocols to grow the cells were described in \cite{segura2021tumor} and \cite{gargini2020idh}. The cells express the luciferase reporter to monitor tumor growth.

Mice were treated with TMZ ($10$ mg/kg/dose) through intraperitoneal injection, with the schedules given in the various experiments. TMZ was dissolved in PBS+1\% BSA, which was used to treat control animals. Different doses of RT were administered to sedated animals with a Multirad225 X-ray irradiator (Faxitron) using a copper filter (0.50 mm) and a Quad Fixture Shield (Precision X-ray) to irradiate only the head of the mice.  Tumor growth was monitored by bioluminescence in an IVIS equipment (Perkin Elmer) after intraperitoneal injection of D-luciferin (75 mg/kg; Thermo Fisher Scientific). Animals were sacrificed when they showed symptoms of disease. The SVZ model was treated with different protocols, varying in both the type and timing of the treatments, and longitudinal data were collected. 


To calibrate the parameters associated with RT, we performed experiments consisting of administering three daily RT sessions of 3 Gy each, starting $39$ days after tumor injection. OS and longitudinal monitoring data of tumor bioluminescence using the In Vivo Imaging System (IVIS) were collected for each mouse. The IVIS is a widely used technique for measuring light emission, particularly in oncology, due to its low cost and non-invasive nature. In the present study, it was employed to assess tumor cells that have been genetically modified to express luciferase, an enzyme that catalyzes a luminescent reaction. The intensity of the bioluminescence signal captured by IVIS is correlated with the number of viable tumor cells expressing luciferase. Previous studies have demonstrated a strong correlation \textit{in vitro} between the number of tumor cells and the intensity of their bioluminescence signal (R$^2=0.99$), as well as between tumor volume and the data captured by the IVIS (R$^2=0.97$) \cite{IVIS}. We directly used bioluminescence data collected for six mice in the following experimental setting: tumors were allowed to grow for $39$ days without treatment, followed by $3$ consecutive days of RT ($3$ Gy/session). IVIS measurements began on day $31$, continuing biweekly until the mice died. Due to its reliability in reflecting tumor burden, all subsequent mouse simulations will use IVIS measurements as a surrogate for tumor size.

For the calibration of the parameters related to TMZ (Section \nameref{sec:TMZ_calibration}), experiment results published elsewhere were used \cite{segura2022optimal}, specifically OS data.

Finally, to validate the model, we performed new experiments with a combined treatment protocol and compared the \textit{in vivo} and \textit{in silico} survivals. We selected a mouse-adapted surrogate of the Stupp protocol \cite{stupp2005radiotherapy}. For these experiments, tumors were allowed to grow for 4 days after cellular injection ($250000$ cells), and then treated with RT (3 Gy/session) and a single concomitant TMZ dose (10 mg/kg) daily for 3 days. After that, there was a 2-day break, and treatment continued with 2 TMZ doses per day (20 mg/kg in total) in three sessions separated by 3 days.

\subsection*{Virtual patients, \textit{in silico} trials, and digital twins}

A primary outcome of the proposed model is its capacity to be personalized using subject-specific data. Once validated, the model can be used to perform analyses, make predictions, optimize treatment strategies, and conduct \textit{in silico} trials.

In Section \nameref{sec:model}, we presented a mathematical model to describe cancer growth and its evolution in response to RT and CT. The model parameters (Table \ref{tab:param}) represent tumor characteristics, including demographic processes (top group of rows), response to RT (middle group), and response to TMZ (bottom group). By assigning specific parameter values, we can generate a virtual patient (VP), which is a digital representation of a patient. When parameter distributions are unknown, as in this study, they are typically assumed to follow a random distribution, as in Ayala et al. work \cite{ayala2021optimal}. Thus, a particular VP is a specific combination of the model parameters in Table \ref{tab:param}, where each parameter is randomly drawn from its range. Repeating this process $N$ times produces a virtual cohort of $N$ VPs. By simulating the responses of VPs to specific protocols, an \textit{in silico} trial is conducted.

By calibrating these parameters to reflect data from a specific subject (Subsection \nameref{subsec:modelparameters}), a digital twin (DT) is generated. This virtual clone replicates the real patient's behaviors under identical conditions \cite{wu2022integrating}. A cohort of $N$ DTs can therefore be created using data from $N$ real subjects. While VP cohorts can be generated in arbitrary sizes, DT cohort sizes are limited by the availability of subject-specific data. Moreover, parameter values estimated from DTs can help refine the parameter distributions used to generate VPs.

\section*{Results}\label{sec:results}

\subsection*{\textit{In vivo} calibration on RT experiments}\label{sec:RT_calibration}

\begin{figure}[h!]
    \centering
    \includegraphics[width=\linewidth]{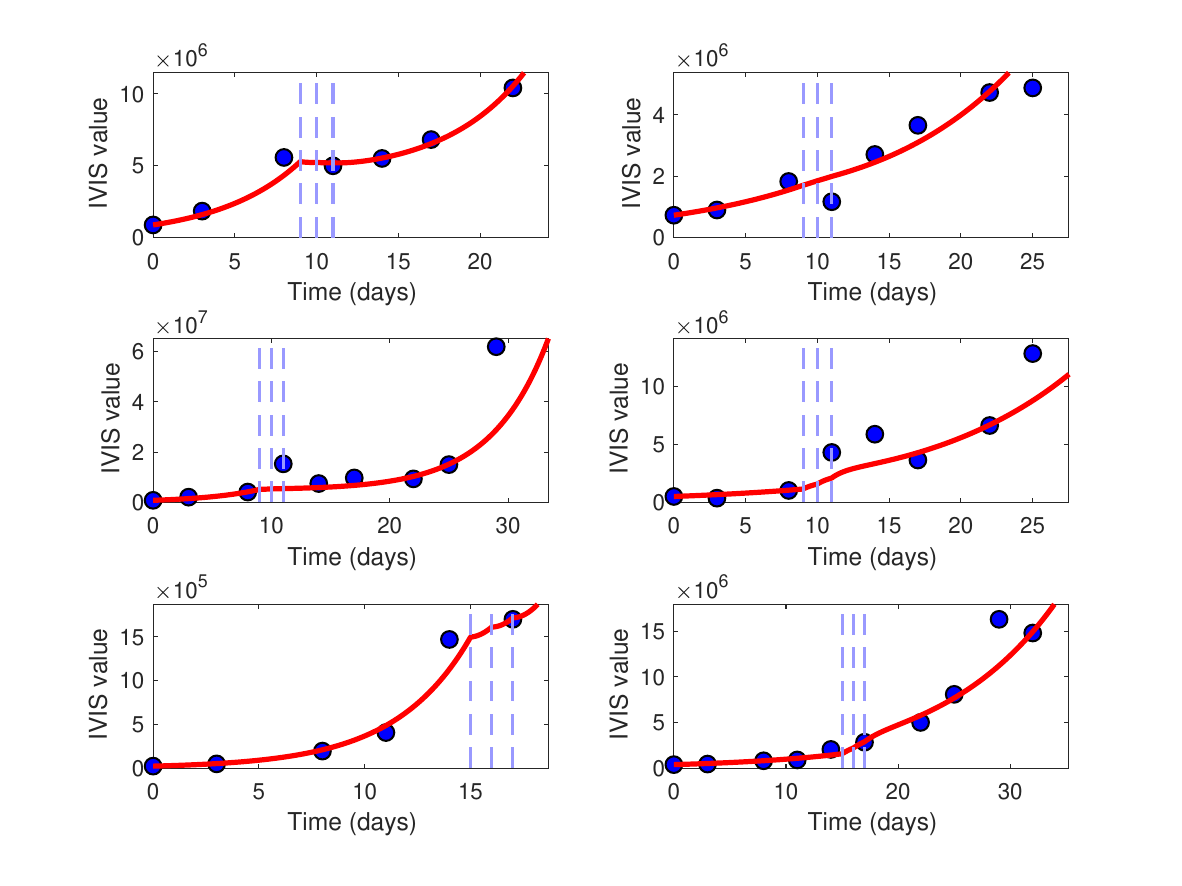}
    \caption{Longitudinal tumor dynamics before and after RT treatment: \textit{In vivo} data (blue points) and numerical simulations (red lines) of model \eqref{eq:systemEDO} and RT effects \eqref{eq:systemRT} equations. Parameters are given in Table \ref{tab:ajusteRT}. Dashed lines represent the times of RT applications. The different subplots correspond to the dynamics of response in six different individuals (mice).}
    \label{fig:ajusteRT}
\end{figure}

In a first set of experiments, mice received RT alone. Data were collected as explained in Section \nameref{subsec:experimentalmethods}.
Tumor size data in mice were obtained using the IVIS  before, during, and after treatment, as shown in Fig \ref{fig:ajusteRT}. 

  \begin{table}[!ht]
  \centering
  \begin{tabular}{c c c c c c c c } 
        Mouse & $\rho_S$ & $\mu_{SQ}$ &$\tau$   &$\gamma$ &$\alpha_{RT}$ & Error\\ \hline
        $1$  &   $1.9913$  & $1.7959$ &$0.7527$ & $0.9259$  &$0.0116$  &  $0.0678$ \\ 
        $2$  &    $0.9433$  & $1.8581$ &$0.0820$ & $0.7485$  &$0.0772$  &  $0.1811$ \\ 
        $3$  &   $2.0745$  & $1.8676$ &$0.0446$ & $0.9011$  &$0.0102$  &  $0.2335$ \\
        $4$  &   $0.9134$  & $1.8182$ &$0.3057$ & $0.5162$  &$0.6206$  & $0.2858$ \\  
        $5$  & $2.7010$  & $2.4658$ &$0.4867$ & $0.7971$  &$0.0529$  &  $0.0733$ \\        
        $6$  &   $0.9514$  & $1.4797$ &$0.5416$ & $0.6184$  &$0.5744$  & $0.1152$ \\ \hline
    \end{tabular}\caption{Model parameters from ODE system governing tumor dynamics (Eqs \eqref{eq:S}-\eqref{eq:Q}-\eqref{eq:D}) and equations modeling RT effect (Eqs \eqref{eq:systemRT}). Parameters are fitted on RT experiments by using the least squares method and mean relative error calculated for each mouse as in Eq \eqref{eq:error}.}\label{tab:ajusteRT}
      \end{table}

The model was calibrated by determining the parameter combination that minimized the mean relative error between real ($T_j$) and simulated data ($T(t_j)$) at time points $t_j$, with $j = 0,1, \dots,N$, specifically: \begin{equation}\label{eq:error}
    \text{Error}=\frac1N\sum_{j=1}^{N}\frac{|T_j-T(t_j)|}{T_j},
\end{equation}using the \texttt{fmincon} Matlab function.  In the simulations shown in Fig \ref{fig:ajusteRT}, the initial total tumor size, $T_0$, corresponds to the first IVIS datum. Initially, the populations of damaged, persister, and resistant cells were assumed to be zero. Thus, the initial tumor was entirely composed of sensitive and quiescent cells. The initial populations of sensitive and quiescent cells were computed using the observed Ki-67 expression and Eqs (\ref{Ki670}).

The fitted parameter values and the corresponding error for each mouse are reported in Table \ref{tab:ajusteRT}. Fig \ref{fig:ajusteRT} presents the collected experimental data alongside the simulated evolution in time of the IVIS for each mouse. Note the close agreement observed between the experimental points and the solid curve obtained from the simulation, showing the model's ability to describe the dynamics observed in the animal model of tumor response to RT.

\subsection*{\textit{In vivo} calibration on TMZ experiments }\label{sec:TMZ_calibration}

In these experiments, mice received TMZ alone following different administration schedules presented elsewhere \cite{segura2022optimal}, with survival time as the only recorded data point per mouse. Given the lack of volumetric longitudinal data, we calibrated only the TMZ response parameter ($\alpha_{E_D}$) at the population level, using the previously estimated model parameters presented in Section \nameref{sec:RT_calibration} and in Table \ref{tab:ajusteRT}. To validate our approach, we replicated the statistical differences in survival times reported in \cite{segura2022optimal} between control and treated mice. This study tested TMZ regimens with three consecutive doses on the same day (4 h apart) and intervals of 1 ($x$+1), 4 ($x$+4), 7 ($x$+7), or 13 days ($x$+13) between cycles, assessing survival differences using the log-rank test.

The three daily TMZ injections were modeled using $\delta_{E_{D_j}}=1/3$ for each TMZ injection, administered $4.8$ hours apart. Note that by assuming a constant sensitive proliferation rate $\rho_S$ (previously fitted for each virtual mouse in Section \nameref{sec:RT_calibration}) and a constant percentage of proliferative cells (Ki-67 value) at the time of tumor injection, it was possible to estimate a tumor IVIS value at injection time for each mouse. 
As before, the untreated tumor was assumed to be composed only of sensitive and quiescent cells. Thus, to calculate initial tumor conditions, we used the first Ki-67 data point as initial total tumor $T_0$ to obtain the sensitive population value at the initial time, $S_0=\text{Ki-}67\cdot T_0,$ and to calculate the sensitive population at the time of tumor injection, $\tilde S_0= S_0{\exp{\left(-\rho_S \cdot \hat t\right)} },$ where $\hat t$ is the time between tumor injection and the first IVIS datum. Finally, $\tilde T_0=\tilde Q_0+\tilde S_0 $ is the initial tumor condition, where $\tilde Q_0 =  \tilde S_0\cdot\left(1-\text{Ki-}67\right)/\text{Ki-}67$ represents the initial quiescent tumor population. 

Taking into account the last IVIS data collected before the mouse death, we estimated a fixed fatal IVIS value of $3\cdot 10^8$.

\begin{figure}
    \centering
    \includegraphics[width=\textwidth]{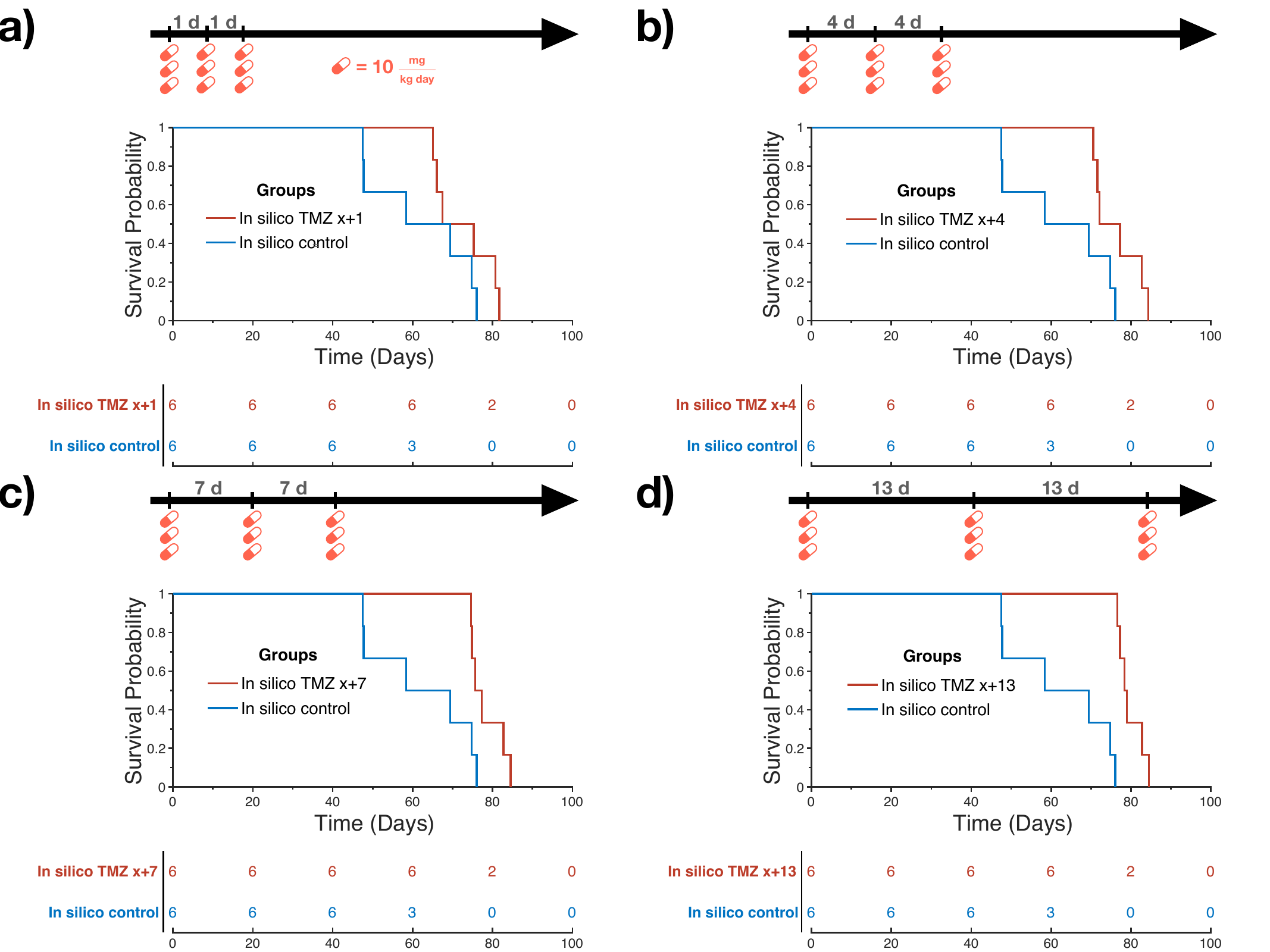}
    \caption{Kaplan-Meier curves and risk tables obtained from \textit{in silico} trials applying TMZ with different scheduling. As in Segura et al. work \cite{segura2022optimal}, we performed TMZ (here \textit{in silico}) trials with three chemotherapy sessions, composed each one of three doses. Chemotherapy sessions were spaced by a) 1 day for protocol x+1, b) 4 days for protocol x+4, c) 7 days for protocol x+7, or d) 13 days for protocol x+13. Model equations \eqref{eq:systemEDO} were used to conduct these \textit{in silico} trials.}
    \label{fig:TMZ_res_cal}
\end{figure}

We found a qualitative statistical agreement with $\alpha_{E_D}=150$. The \textit{in silico} experiments are summarized in Figure 
\ref{fig:TMZ_res_cal}. The log-rank test revealed significant differences between the control and treated groups for the TMZ $x$+13 (p $= 0.002$) and TMZ $x$+7 (p $=0.049$) protocols. However, no significant differences were observed for the TMZ $x$+4 (p $=0.105$) and TMZ $x$+1 (p $=0.318$) protocols. The results are in line with experimental data \cite{segura2022optimal}, reporting no significant differences for the TMZ $x$+1 and $x$+4 protocols, and $p$-values of 0.03 and 0.011 for the TMZ $x$+7 and $x$+13 protocols, respectively. For more details, we refer the reader to the \textit{in vivo} results in Segura et al. work \cite{segura2022optimal}.

\subsection*{\textit{In vivo} validation on combined treatment experiments}\label{sec:QRT_DT_val}

\begin{figure}[ht]
\centering
    \includegraphics[width=\textwidth]{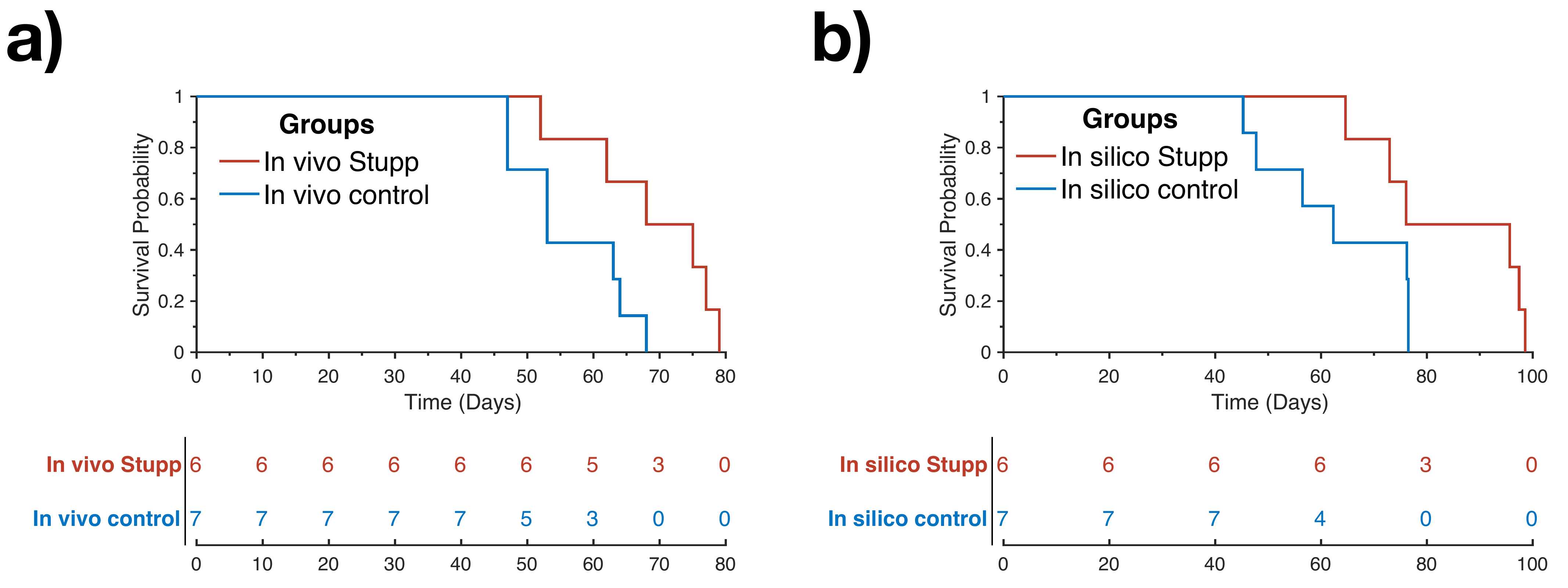}
\caption{Kaplan-Meier curves and risk tables obtained from \textit{in vivo} and \textit{in silico} trials pplying the Stupp protocol. A \texttt{log-rank} test in Matlab revealed no statistically significant differences between the control and treated (Stupp) groups neither in the \textit{in vivo} (p $= 0.092$) nor \textit{in silico} (p $= 0.105$) experiments. For these experiments, treated mice underwent three consecutive daily radiotherapy sessions (3 Gy per session) with concomitant TMZ (10 mg/kg/day). After a two-day break, they received three additional adjuvant TMZ sessions (20 mg/kg/day), administered every three days. Model \eqref{eq:systemEDO} and RT effects \eqref{eq:systemRT} equations were used to conduct these \textit{in silico} trials.}
\label{fig:sim_OSstupp_vivo_silico}
\end{figure}

To validate the model, we compared real and simulated survival times under the same treatment conditions. Since tumor growth and treatment response parameters were calibrated in previous experiments, we focused on replicating the statistical differences in survival distributions between control and treated mice.
These experiments involved RT and TMZ following a mouse-adapted surrogate of the Stupp protocol for humans (Fig \ref{fig:sim_OSstupp_vivo_silico}). Thus, treatment started with concomitant RT and TMZ, followed by adjuvant TMZ at double the dose administered during concomitant RT and TMZ. Initial conditions and fatal size considered are those mentioned in Section \nameref{sec:TMZ_calibration}. Each TMZ administration was simulated with $\delta_{E_{D_j}} = 1/3,\,j=1,2,\dots,9$, with 4.8-hour intervals between consecutive injections administered on the same day. On RT+TMZ days, RT was given first, followed by TMZ, as in the \textit{in vivo} experiments.

The \textit{in vivo} and \textit{in silico} results are reported in Fig \ref{fig:sim_OSstupp_vivo_silico}, done with \cite{creed2020matsurv} showing the resulting KM curves, and the risk tables (number of alive mice over time).

\subsection*{\textit{In silico} optimal treatment}\label{sec:optimaltreatment}

Once the model was validated, we used it to investigate the optimal treatment for the calibrated virtual cohort (six mice in total) at the population level through \textit{in silico} clinical trials. 
We considered the patient-specific (mouse-specific) parameters shown in Table \ref{tab:ajusteRT}, along with the other parameters from Table \ref{tab:param} and $\alpha_{E_D}=150$. The initial tumor conditions and fatal size were the same as in Section \nameref{sec:TMZ_calibration}. 

Mathematically, this is a control problem where the objective function is to maximize the median OS, and the control variables are the RT and CT administrations. 
Note that the solution space of all possible treatments is infinite-dimensional.
To reduce complexity and mimic both the \textit{in vivo} experiments used to calibrate and validate the model but also real clinical settings, we limited our search by fixing both the number of doses and their dosages of both RT ($3$ Gy per session, $3$ sessions) and TMZ ($10$ mg/kg per injection, three injections per session, three sessions). Thus, we modeled TMZ administration using $\delta_{E_{D_j}} = 1/3$ for each injection $j=1,\dots,9$, and separating consecutive injections administered in the same day by $4.8$ hours, as in Section \nameref{sec:TMZ_calibration}. If RT and TMZ are administered on the same day, RT is given first, followed by the first TMZ injection, as in Section \nameref{sec:QRT_DT_val}.  
In addition, we also fixed the intervals between consecutive RT and TMZ sessions (which were equally spaced) to obtain a realistic optimal schedule consistent with real administration protocols. Considering the observed survival times in mice, we constrained the maximum interval between consecutive RT sessions to $15$ days, the maximum interval between consecutive TMZ sessions to $15$ days, and the maximum delay between the two treatment onsets to $30$ days. Note that treatments began on day $1$ in the simulations.
 
\begin{figure}[h!]
    \centering
    \includegraphics[width=\linewidth]{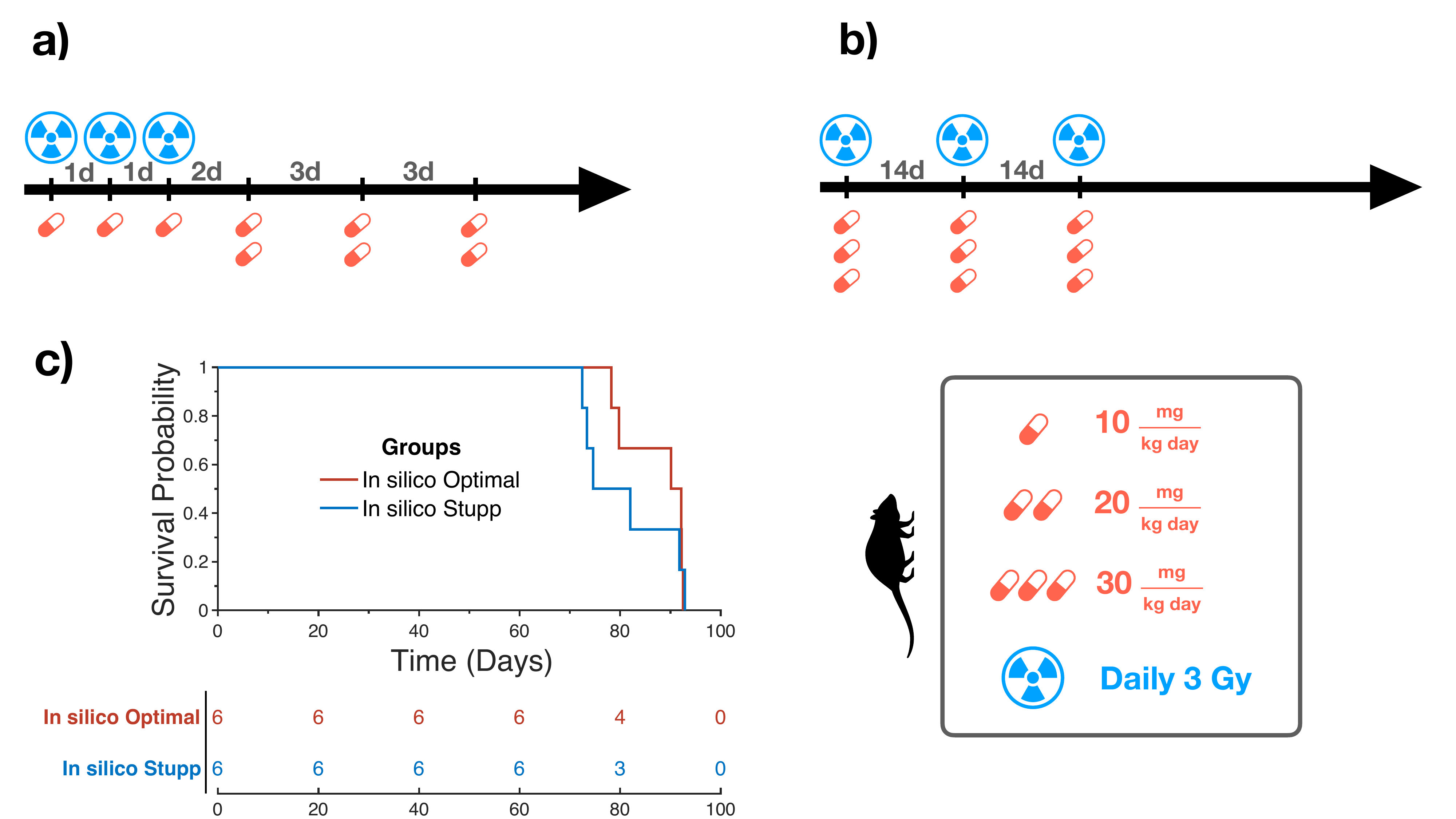}
    \caption{a) Stupp protocol in mice. b) Optimal treatment protocol in mice. c) Kaplan-Meier curves and risk tables obtained from \textit{in vivo} and \textit{in silico} trials applying the Stupp and optimal treatment protocols.}
    \label{fig:mice_opt_vs_stupp}
\end{figure}

Under these conditions, the highest median OS for the virtual cohort was achieved with a concomitant schedule, where RT and TMZ were administered on the same day, with a $15$-day interval between each chemoradiotherapy session. Specifically, the identified optimal treatment achieved a median OS of $91.13$ days, representing a gain of 16.35\% (12.8 days) compared to the Stupp protocol, which had a median OS of 78.33 days. The treatment schemes of the Stupp and the optimal protocols are presented in Fig \ref{fig:mice_opt_vs_stupp}-a) and \ref{fig:mice_opt_vs_stupp}-b), respectively. The \textit{in silico} trial results comparing the Stupp protocol and the identified optimal treatment are reported in Fig \ref{fig:mice_opt_vs_stupp}-c).


\subsection*{Scaling to human patients} \label{sec:modelhumanpatients}

Our goal was to improve the standard Stupp treatment and to identify optimal treatments for humans. Therefore, we scaled our \textit{in vivo} results to humans. First, we adapted the calibrated \textit{in vivo} model from mice to humans. Then, translated the identified \textit{in vivo} optimal treatments to humans and real clinical settings. Clearly, the optimal treatment should outperform the standard Stupp protocol, that is the gold standard of treatment.

To adapt our calibrated \textit{in vivo} model for humans, we retained the same design used for mice in Section \nameref{sec:model} (Fig \ref{fig:model}). However, we adjusted the parameter distributions to reflect human ranges. Table \ref{tab:paramhumans} summarizes the model parametrization for humans. 

\begin{table}[!ht]
\centering 
\caption{Summary of the model parameters in humans, including symbol, range, unit, meaning, and sources. Note that the parameters in the top, middle, and bottom row groups represent demographic processes, response to radiotherapy (RT), and response to temozolomide (TMZ), respectively. Some parameter ranges are constrained to their feasible values (FV).}
\begin{tabular}{ccccc}
\textbf{Par.} & \textbf{Meaning} & \textbf{Range} & \textbf{Unit} & \textbf{Source}\\ \hline
{$\rho_S$} & \makecell{Sensitive population \\ proliferation rate} & {$[10^{-3},10^{-1}]$} & {d$^{-1}$} & \cite{Ode2}\\ 
{$\rho_R$} & \makecell{Drug-resistant population \\ proliferation rate} & {$[10^{-3},10^{-1}]$} & {d$^{-1}$} & \cite{gupta2014discordant} \\
$\mu_{SQ}$ & \makecell{Mutation rate from \\ sensitive to quiescent} & $[0.2,0.6]$ & d$^{-1}$ & \cite{segura2022optimal} \\ 
$\mu_{QS}$ & \makecell{Mutation rate from \\ quiescent to sensitive} & $[0.03,0.2]$ & d$^{-1}$ & \cite{segura2022optimal} \\ 
$\mu_{PS}$ & \makecell{Mutation rate from \\ persister to sensitive} & $0.1$ & d$^{-1}$ & \cite{delobel2023overcoming} \\
Ki-67 & \makecell{Fraction of \\ proliferating cells} & $[0.1,0.5]$ & - & \cite{delobel2023overcoming}\\\hline
$\gamma$ & \makecell{Proliferation induced in \\ quiescent population by RT} & $[0,1]$ & - & FV \\ 
$\alpha_{RT}$ & \makecell{Fraction of sensitive \\ population damaged by RT} & $[0,1]$ & - & FV\\ 
$\tilde\alpha_{RT}$ & \makecell{Fraction of drug-resistant \\ population damaged by RT} & $[0,1]$ & - & FV \\ 
$\tau$ & \makecell{Damaged population \\ elimination rate} & $[\frac{\rho_S}{100},2\rho_S]$ & d$^{-1}$ & \cite{joiner2025basic}\\\hline
$\beta_{SP_I}$ & \makecell{Persister cells \\ generation rate} & $0.3560$ & d$^{-1}$ & \cite{delobel2023overcoming} \\ 
$\beta_{PR}$ & \makecell{Resistant cells \\ generation rate} & $0.0697$ & d$^{-1}$ & \cite{delobel2023overcoming} \\ 
$\alpha_{E_D}$ & \makecell{TMZ killing efficiency} & $0.4469$ & d$^{-1}$ & \cite{delobel2023overcoming} \\ 
$\lambda$ & \makecell{TMZ efficacy \\ decay rate} & $8.3178$ & d$^{-1}$ & \cite{rudek2004temozolomide} \\ 
$\delta_{E_{D_j}}$ & \makecell{TMZ efficacy increment \\ of dose $j$} & $[0,1]$ & - & FV \\ \hline
\end{tabular}
\label{tab:paramhumans}
\end{table}

To ensure clinical feasibility, we fixed the radiation doses at $2$ Gy per irradiation, with one irradiation per day. RT was considered to be administered only on working days (from Monday to Friday). TMZ, administered orally, was considered to be given at daily doses of $75$ mg/m$^2$/day when concomitant with RT, or $150$ mg/m$^2$/day when adjuvant to RT.

In summary, we simulated schedules with $6$ chemoradiotherapy sessions as in the concomitant part of the Stupp protocol (irradiation on Mondays to Fridays and $75$ mg/m$^2$/day of TMZ daily) separated $1,2,\dots,8$ weeks. However, considering the potential for reduced toxicity with treatment breaks and extended intervals \cite{jackson2019mid}, we also explored the possibility of administering additional cycles, up to a total of $12$ chemoradiotherapy sessions.

\begin{figure}
    \centering
        \includegraphics[width=\textwidth]{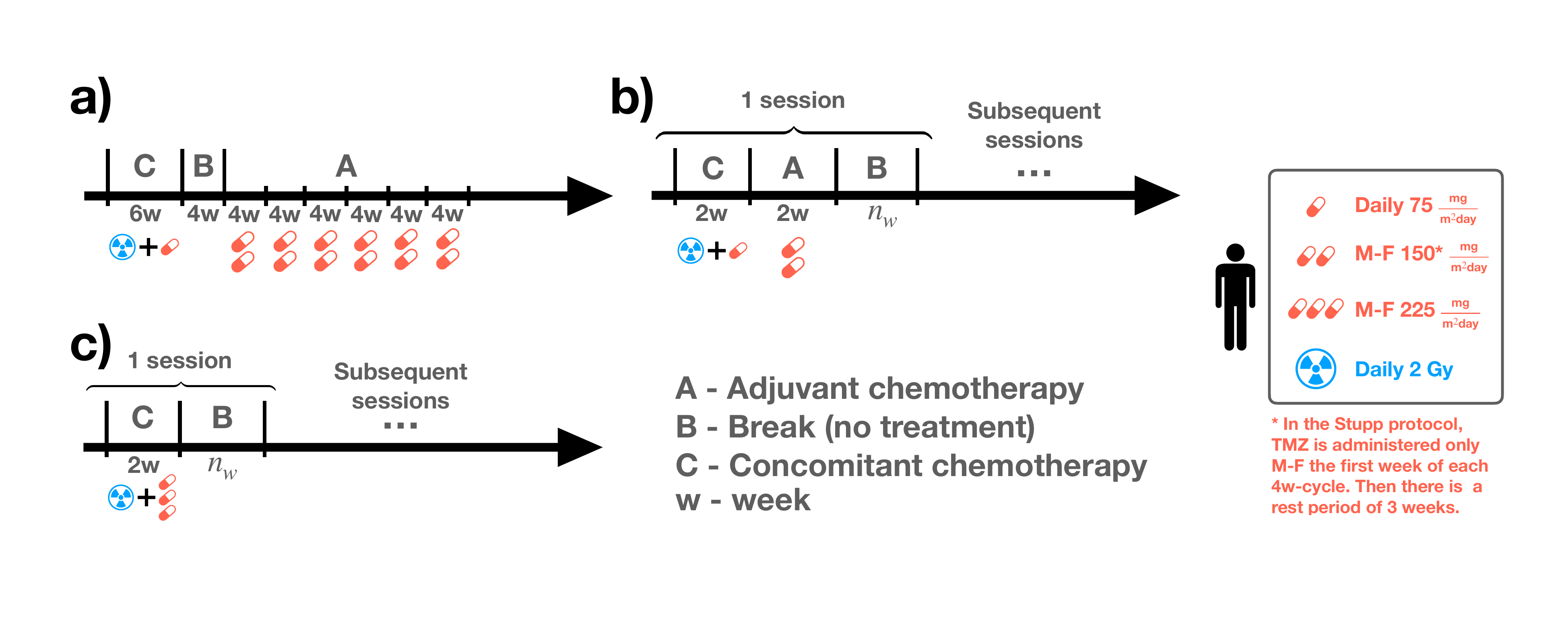}
    \caption{Graphical representation of key protocols: a) Stupp protocol in humans, b)-c) optimal protocols in humans deduced from optimal protocol in mice.}
    \label{fig:protocols}
\end{figure}

Figs \ref{fig:mice_opt_vs_stupp} and \ref{fig:protocols} present the key administration protocols investigated in this study: Stupp protocol in mice(Fig \ref{fig:mice_opt_vs_stupp}-a)), optimal protocol in mice (Fig \ref{fig:mice_opt_vs_stupp}-b)), Stupp protocol in humans (Fig \ref{fig:protocols}-a)), and optimal protocol in humans deduced from the optimal protocol in mice (Fig \ref{fig:protocols}-b) and \ref{fig:protocols}-c)). 
Specifically, we assumed that a RT and a CT session in mice corresponded to a cycle of RT and CT in humans, respectively. Moreover, while mice received in total 3 RT sessions in the experiments, humans underwent 6 weeks of total (concomitant) RT within the Stupp protocol. Thus, 1 RT session \textit{in vivo} could be translated into 2 weeks of RT in humans. Based on the differences between the Stupp protocol and the optimal treatment identified in mice, we applied the corresponding modifications to the Stupp protocol in humans. This led us to two distinct strategies. First, we explored concomitant chemoradiotherapy sessions separated by treatment breaks, meaning that we administered concomitant RT and TMZ for two weeks, followed by a pause of $n_w$ weeks, where we investigated $n_w=1,2,\dots,8$ weeks (Fig \ref{fig:protocols}-b)). Second, we investigated concomitant chemoradiotherapy plus adjuvant CT sessions, separated by treatment breaks. In this approach, we administered concomitant RT and TMZ for one or two weeks, followed by adjuvant TMZ for one or two weeks, and then a break of $n_w$ weeks, where we investigated $n_w=1,2,\dots,8$ weeks (Fig \ref{fig:protocols}-c)). 
Protocols corresponding to concomitant (C) and adjuvant (A) sessions followed by $n_w$ weeks of break (B) are denoted as CA$n_w$B, while those corresponding to concomitant (C) sessions followed by $n_w$ weeks of break (B) are denoted C$n_w$B, with $n_w=1,2,\dots,8$.

To demonstrate the superiority of the translated optimal treatment in humans, derived from the \textit{in silico} optimal treatment for mice in Section \nameref{sec:optimaltreatment}, we conducted various \textit{in silico} trials with VPs ($N=1000$), following the methodology described in Section \nameref{sec:results}. We used an initial tumor volume of 50 cm$^3$ and a fatal tumor volume of 280 cm$^3$ \cite{delobel2023overcoming}. Thus, the survival time of a VP corresponds to the time its tumor required to reach the fatal volume.

\begin{figure}
    \centering
    \includegraphics[width=\linewidth]{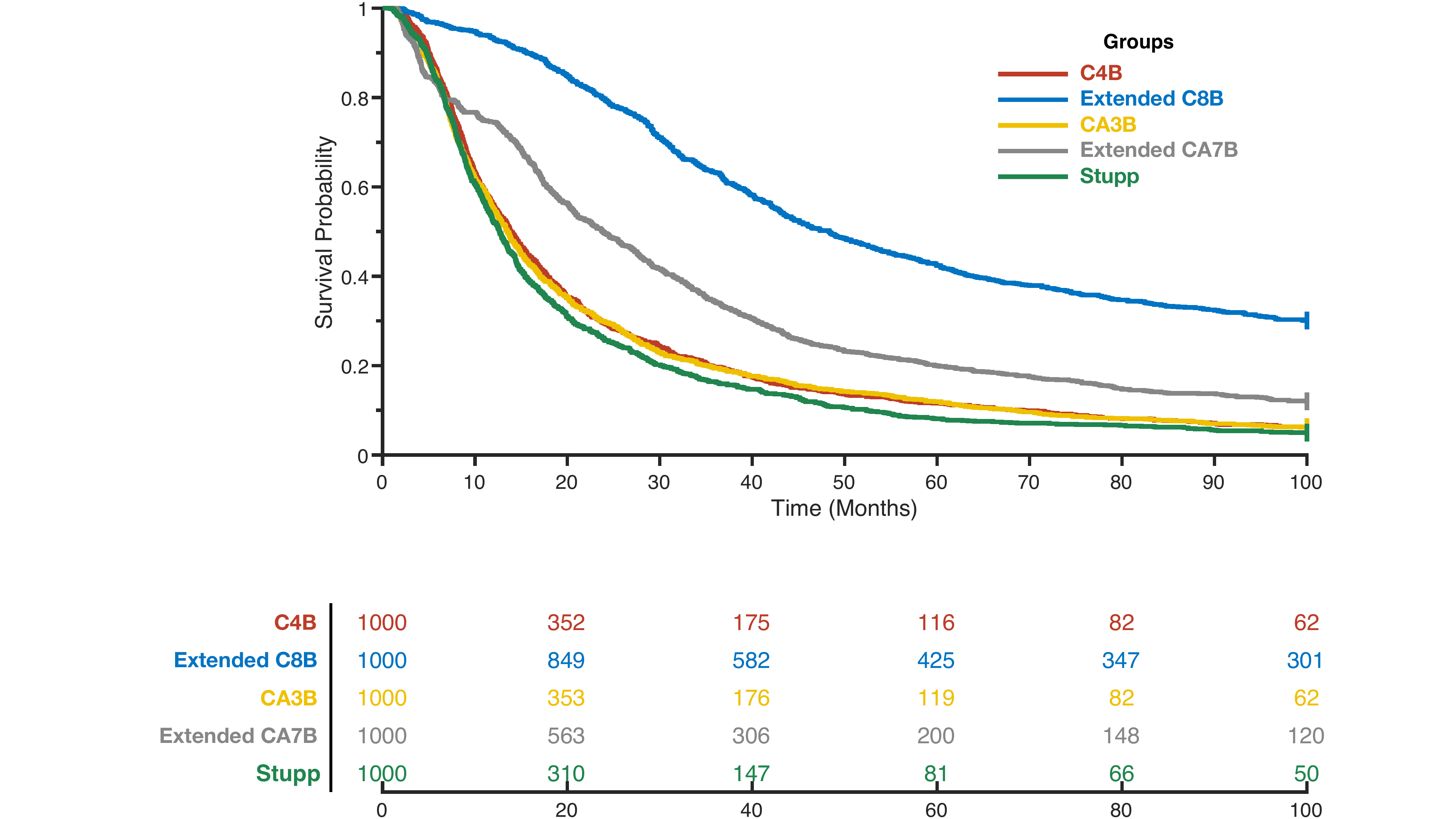}
    \caption{Kaplan-Meier curve and risk table comparing the \textit{in silico} trial in humans treated with the standard Stupp protocol and the alternative proposed protocols with and without extended administrations. C4B treatment consists of three sessions composed of 2 weeks of concomitant (C) RT and CT followed by 4 weeks of break (B), i.e., Fig \ref{fig:protocols}-b) with $2$ subsequent sessions and $n_w=4$. Similarly, extended C8B treatment follows the same structure but with 8 weeks of break and extends to six sessions in total.
   CA3B treatment comprises three sessions, each divided into a 2 weeks of concomitant (C) RT and CT, followed by 2 weeks of adjuvant (A) CT and then by 3 weeks of break (B), i.e., Fig \ref{fig:protocols}-c) with $2$ subsequent sessions and $n_w=3$. Extended CA7B treatment follows the same structure but with 7 weeks of break and extends to six sessions in total.}
    \label{fig:KM_opt_ttos}
\end{figure}

The OS results from the \textit{in silico} trials are presented in Fig \ref{fig:KM_opt_ttos}.
The median OS achieved with the Stupp protocol was $12.66$ months, which was surpassed by the protracted administration protocols. Specifically, the best outcome for 
C$n_w$B protocols occurred with $n_w=4$ (1-month break), yielding a median OS of $13.86$ months, representing a 1.2-month increase (9.42 \%).
The best outcome for CA$n_w$B protocols was achieved with $n_w=3$ (three-week break), obtaining a median OS of $13.52$ months, that is, a 0.86-month gain (6.77 \%).

The proposed protracted protocols are less intensive and lead to lower toxicity for patients \cite{jackson2019mid}. Thus, we assumed that patients could receive doubled sessions of RT and TMZ. Consequently, we doubled 
C$n_w$B and CA$n_w$B protocols. These new treatment schedules are referred to as extended C$n_w$B and extended CA$n_w$B protocols, respectively. The resulting survival rates improved considerably. 
The best result for C$n_w$B was achieved with $n_w=8$ (2 months of treatment break), yielding a median OS of $48.35$ months, corresponding to a gain of $35.69$ month (281.82 \%).
The best result for CA$n_w$B was achieved with $n_w=7$ (7 weeks of treatment break), obtaining a median OS of $23.84$ months, corresponding to a gain of $11.18$ months (88.29 \%).

\section*{Discussion}

In this work, we investigated the response of MGs to combined chemoradiotherapy using a mathematical framework that incorporated modeling, experimental calibration and \textit{in silico} trials. The proposed model integrates key biological mechanisms into a system of ODEs, effectively translating known biological insights into a computationally manageable framework. We considered the role of Ki-67 as a marker of cell proliferation \cite{sun2018ki}, cancer cell dormancy \cite{dormantstateandback}, resistance to TMZ via persister cell pathways \cite{rabe2020identification}, and the interactions between RT and TMZ effects on tumor cell populations \cite{lomax2013biological,agarwala2000temozolomide}. By extending our findings from an \textit{in vivo} mouse model to a human context, we proposed new strategies that could enhance patient survival compared to the standard Stupp protocol \cite{stupp2005radiotherapy}. Specifically, we explored the use of protracted concomitant CT (C$n_w$B protocols) and concomitant chemoradiotherapy followed by adjuvant CT (CA$n_w$B protocols), both of which demonstrated significant advantages in terms of survival while offering a reduction in treatment intensity and toxicity.

Our approach builds upon a series of prior studies that utilize ODE-based models to describe tumor growth and treatment responses through \textit{in silico} trials. Various ODE models have been employed to investigate MG progression under treatments with TMZ, RT, and their combinations, providing valuable insights into dose scheduling strategies, resistance mechanisms, and treatment efficacy. For instance, Victoria et al. explored the response to RT \cite{victoria2015incorporating}, proposing an optimal scheduling that involves less intensive, fractionated RT. Regarding TMZ, Panetta et al. described the drug's toxicity \cite{panetta2003mechanistic}, while Pérez et al. recently presented a model calibrated with data from GBM spheroids, examining both TMZ-sensitive and TMZ-resistant spheroids using a compartmental approach \cite{perez2024modelling}. More complex models have also incorporated the role of the vasculature in treatment response \cite{yan20173d}. Additionally, the influence of pH as a regulator of TMZ's anti-tumor efficacy has been highlighted in recent works \cite{tafech2023characterization, stephanou2019ph}.

Other studies have proposed optimal TMZ scheduling strategies using hybrid discrete-continuous models \cite{oraiopoulou2024temozolomide}, agent-based models \cite{surendran2023agent}, and artificial intelligence approaches \cite{houy2018optimal, zade2020reinforcement}, although these approaches primarily focus on short-term processes and do not account for resistance mechanisms. In contrast, Sorribes et al. proposed an optimal combination of TMZ and an MGMT inhibitor based on a mathematical model that explicitly considers resistance to TMZ \cite{sorribes2020mitigating}.

TMZ resistance  was incorporated into an ODE model describing tumor dynamics under TMZ treatment, calibrated with human data and used to propose an alternative administration schedule, with extended intervals between consecutive TMZ administrations \cite{delobel2023overcoming}. The concept was validated in experiments on animal models \cite{segura2022optimal}.
Here, we extend previous models that considered only TMZ resistance by incorporating the effects of RT and the synergies between these two treatments. In addition, our work is focused on the development of a robust validated mathematical model able to capture the complex biological dynamics in MGs, underlying tumor progression and therapy resistance in long-term treatment scenarios. These efforts align with and extend the broader literature, which demonstrates the efficacy of ODE-based frameworks for studying MGs, GBMs, and other aggressive tumors under various treatment regimens \cite{falco2021silico}.

We calibrated an ODE model that describes the main biological mechanisms of MG growth and response to combined RT and TMZ treatment in \textit{in vivo} mouse models. Our approach followed a typical workflow \cite{brady2019mathematical}. First, we identified the key biological processes and, using the best knowledge on those processes, developed the ODE model. Next, the model was calibrated using animal model data and validated with other new data. Finally, we used the model to predict the performance of both known and new treatments.

The optimal treatment identified in the \textit{in silico} trials for the virtual mouse cohort, the protracted concomitant chemoradiotherapy with a 15-day break between treatment sessions, resulted in a 17\% improvement in median OS compared to the Stupp protocol. Although improvements are small, it is important to emphasize that the proposed protocol would have substantially lower toxicity, thus leading to both better survival and less toxicity. This shows that the traditional approach, while effective, could be further optimized by adjusting the timing and frequency of treatments. Indeed, our findings are consistent with previous studies suggesting that optimizing treatment schedules, rather than merely increasing doses, can enhance treatment efficacy and patient outcomes \cite{delobel2023overcoming,ayala2021optimal,segura2022optimal,italia2022optimal,italia2023mathematical}.

Our goal was to investigate optimal treatments for humans. When translating the optimal treatment schedule found to humans, we applied realistic constraints such as using the same radiation and TMZ daily doses that are used currently. Our proposed protracted protocols, especially those extending the treatment duration with periodic breaks, outperformed the standard Stupp protocol in terms of median OS ($12.7$ months for humans in our simulations). Specifically, the protracted concomitant chemoradiotherapy protocol with a 1-month treatment break resulted in a median OS of $13.9$ months, an improvement of 9\% compared to traditional administration schedules, while the extended protocols with doubled sessions separated by 2 months of treatment break resulted in even more substantial improvements in OS, with gains as high as 280\%, achieving a median OS of nearly 4 years ($48$ months). These results underscore the importance of scheduling in optimizing therapeutic outcomes and reducing toxicity associated with intensive treatment regimens.

In addition to providing increased survival and reduced treatment-related morbidity, protracted treatment schedules in humans would offer the significant advantage of reducing toxicity. This reduction likely facilitates the administration of a larger total number of treatment sessions. By spacing out sessions and allowing recovery between rounds, these schedules align with clinical observations that treatment breaks can mitigate side effects and enhance the patient's quality of life \cite{jackson2019mid}. 

The utility of our work extends beyond MGs, as the developed model could be adapted to other cancers exhibiting similar biology, such as EGFR mutant non-small cell lung cancer treated with osimertinib \cite{oren2021cycling}, cisplantin, erlotinib, gefitinib drugs, or EGFR TKI inhibitor \cite{paramgeneracionpersisters}, HER2-amplified breast cancer treated with lapatinib \cite{hangauer2017drug}, and acute myeloid leukemia treated with cytarabine \cite{boyd2018identification, ho2016evolution}, where persister cells have been shown to rule the development of resistances. The ability to recalibrate the model for other malignancies with similar mechanisms of resistance suggests its broader applicability across various tumor types.

One of the primary advantages of using mathematical models in oncology is the ability to perform \textit{in silico} trials \cite{wang2024virtual}. These virtual trials enable the rapid testing of different treatment protocols, offering a safe, cost-effective, and ethical alternative to conventional clinical approaches. Our study highlights how \textit{in silico} trials can accelerate the development of optimized treatment regimens, providing insights into how specific changes in therapy can improve patient outcomes. \textit{In silico} trials can play a crucial role in the design of clinical studies, ensuring that experimental efforts are more efficient and focused on the most promising approaches \cite{brown2022derisking}. Although our results are promising, they must be validated in clinical settings.

Unlike Ki-67, methylation of methylguanine-DNA methyltransferase (MGMT), a key marker of TMZ resistance due to its role in regulating MGMT expression \cite{van2010absence}, has not been explicitly considered in this study. However, Segura et al. \cite{segura2022optimal} found that, in both \textit{in vitro} and \textit{in vivo} experiments, MGMT expression in SVZ models treated with TMZ increases following intensive schedules ($x$+1 and $x$+4, presented in Section \nameref{sec:TMZ_calibration}), whereas it did not under protracted regimens ($x$+7 and $x$+13, Section \nameref{sec:TMZ_calibration}). Their finding provides a rationale for further supporting the protracted protocols proposed based on our model simulation results.

Despite the strengths of ODE-based models--such as computational efficiency and ease of interpretation--they have inherent limitations. For instance, ODE models cannot fully capture the spatial heterogeneity within tumors or the microenvironmental variability that influences treatment outcomes. In particular, while hypoxia has been recognized as a critical factor in GBM progression \cite{martinez2012hypoxic,celora2024characterising,spinicci2025modelling}, it is not explicitly included in our model. 
These factors are essential, as they contribute to differential therapeutic responses across heterogeneous tumor regions. Moreover, the genetic and molecular heterogeneity of tumors, which significantly influences treatment efficacy, is also not accounted for in our model. Addressing these limitations represents a key direction for future work, particularly in the development of more personalized treatment strategies.

In addition, while our model and simulations provide valuable insights, we would like to highlight several limitations. First, while we calibrated the model using data from mice, further validation in larger animal models and human trials is essential to confirm the clinical relevance of our findings. The translation from mice to humans involves a range of assumptions, including dose adjustments and the scaling of treatment cycles, which may not fully capture the complexities of human physiology and tumor behavior. Additionally, the model assumes that all patients respond uniformly to TMZ, but tumor heterogeneity and patient individual differences, such as drug metabolism and response, could significantly affect treatment outcomes. Future studies could incorporate patient-specific parameters to investigate more personalized treatment strategies.

The findings of this study suggest that the proposed protracted chemoradiotherapy protocols could improve OS, opening a new area for clinical investigation.
By incorporating new experimental results into the model, we can further validate or enhance its accuracy and therapeutic predictions. The integration of computational models and experimental data is essential to providing a comprehensive understanding of tumor dynamics and improving clinical strategies. However, tailored clinical trials are essential to fully validate and demonstrate the superiority of the protracted protocol over the standard Stupp protocol and to determine its applicability in real-world settings.

\section*{Conclusion}

This work demonstrates the potential of mathematical models, particularly ODE-based frameworks, and \textit{in silico} trials, to improve MG treatment through optimized chemoradiotherapy regimens. Our simulations suggest that protracted treatment schedules could significantly enhance patient survival and quality of life, offering new strategies for clinical investigation. The protracted chemoradiotherapy schedules proposed in this study provide a potential pathway to refine current clinical practices, offering a more effective and personalized approach to treating these aggressive and often fatal MGs. However, further experimental work is needed to verify the concepts and validate them in clinical settings. We hope this research may contribute to the global activity working towards integrating advanced computational approaches with experimental and clinical data, leading ultimately to the real-world implementation of mathematical models and \textit{in silico} trials in the fight against MGs and other cancers. 

\section*{Declarations}

\subsection*{Conflicts of interest} We declare that we have no competing interests.

\nolinenumbers

%
%
%


\begin{thebibliography}{10}

\bibitem{santucci2020progress}
Santucci C, Carioli G, Bertuccio P, Malvezzi M, Pastorino U, Boffetta P, Negri E, Bosetti C, La Vecchia C
\newblock Progress in cancer mortality, incidence, and survival: a global overview.
\newblock European Journal of Cancer Prevention. 2020;29(5):367--381.
\bibitem{stupp2005radiotherapy}
Stupp R, Mason WP, Van Den Bent MJ, Weller M, Fisher B, Taphoorn MJB, Belanger K, Brandes AA, Marosi C, Bogdahn U, et al.
\newblock Radiotherapy plus concomitant and adjuvant temozolomide for glioblastoma.
\newblock New England Journal of Medicine. 2005;352(10):987--996.

\bibitem{agarwala2000temozolomide}
Agarwala SS, Kirkwood JM
\newblock Temozolomide, a novel alkylating agent with activity in the central nervous system, may improve the treatment of advanced metastatic melanoma.
\newblock The Oncologist. 2000;5(2):144--151.

\bibitem{roos2007apoptosis}
Roos WP, Batista LFZ, Naumann SC, Wick W, Weller M, Menck CFM, Kaina B
\newblock Apoptosis in malignant glioma cells triggered by the temozolomide-induced {DNA} lesion {O6}-methylguanine.
\newblock Oncogene. 2007;26(2):186--197.

\bibitem{rudek2004temozolomide}
Rudek MA, Donehower RC, Statkevich P, Batra VK, Cutler DL, Baker SD
\newblock Temozolomide in patients with advanced cancer: phase I and pharmacokinetic study.
\newblock Pharmacotherapy: The Journal of Human Pharmacology and Drug Therapy. 2004;24(1):16--25.

\bibitem{wick2009new}
Wick W, Platten M, Weller M
\newblock New (alternative) temozolomide regimens for the treatment of glioma.
\newblock Neuro-oncology. 2009;11(1):69--79.

\bibitem{singh2021mechanisms}
Singh N, Miner A, Hennis L, Mittal S
\newblock Mechanisms of temozolomide resistance in glioblastoma—a comprehensive review.
\newblock Cancer Drug Resistance. 2021;4(1):17.

\bibitem{sachs2016optimal}
Sachs JR, Mayawala K, Gadamsetty S, Kang SP, de Alwis DP
\newblock Optimal dosing for targeted therapies in oncology: drug development cases leading by example.
\newblock Clinical Cancer Research. 2016;22(6):1318--1324.

\bibitem{italia2023mathematical}
Italia M, Wertheim KY, Taschner-Mandl S, Walker D, Dercole F
\newblock Mathematical model of clonal evolution proposes a personalised multi-modal therapy for high-risk neuroblastoma.
\newblock Cancers. 2023;15(7):1986.
\bibitem{arias2017metabolomics}
Arias-Ramos N, Ferrer-Font L, Lope-Piedrafita S, Mocioiu V, Juli{\`a}-Sap{\'e} M, Pumarola M, Ar{\'u}s C, Candiota AP
\newblock Metabolomics of therapy response in preclinical glioblastoma: {A} multi-slice {MRSI}-based volumetric analysis for noninvasive assessment of temozolomide treatment.
\newblock Metabolites. 2017;7(2):20.

\bibitem{ferrer2017metronomic}
Ferrer-Font L, Arias-Ramos N, Lope-Piedrafita S, Juli{\`a}-Sap{\'e} M, Pumarola M, Ar{\'u}s C, Candiota AP
\newblock Metronomic treatment in immunocompetent preclinical {GL}261 glioblastoma: effects of cyclophosphamide and temozolomide.
\newblock NMR in Biomedicine. 2017;30(9):e3748.

\bibitem{calero2021immune}
Calero-P{\'e}rez P, Wu S, Ar{\'u}s C, Candiota AP
\newblock Immune system-related changes in preclinical {GL}261 glioblastoma under {TMZ} treatment: {E}xplaining {MRSI}-based nosological imaging findings with {RT}-{PCR} analyses.
\newblock Cancers. 2021;13(11):2663.

\bibitem{wu2020anti}
Wu S, Calero-P{\'e}rez P, Villama{\~n}an L, Arias-Ramos N, Pumarola M, Ortega-Martorell S, Juli{\`a}-Sap{\'e} M, Ar{\'u}s C, Candiota AP
\newblock Anti-tumour immune response in {GL}261 glioblastoma generated by {T}emozolomide {I}mmune-{E}nhancing {M}etronomic {S}chedule monitored with {MRSI}-based nosological images.
\newblock NMR in Biomedicine. 2020;33(4):e4229.

\bibitem{segura2022optimal}
Segura-Collar B, Jim{\'e}nez-S{\'a}nchez J, Gargini R, Dragoj M, Sep{\'u}lveda-S{\'a}nchez JM, Pe{\v{s}}i{\'c} M, Ram{\'\i}rez MA, Ayala-Hern{\'a}ndez LE, S{\'a}nchez-G{\'o}mez P, P{\'e}rez-Garc{\'\i}a VM
\newblock On optimal temozolomide scheduling for slowly growing glioblastomas.
\newblock Neuro-Oncology Advances. 2022;4(1):vdac155.

\bibitem{pasquier2010metronomic}
Pasquier E, Kavallaris M, André N
\newblock Metronomic chemotherapy: new rationale for new directions.
\newblock Nature Reviews Clinical Oncology. 2010;7(8):455--465.

\bibitem{faivre2013mathematical}
Faivre C, Barbolosi D, Pasquier E, André N
\newblock A mathematical model for the administration of temozolomide: comparative analysis of conventional and metronomic chemotherapy regimens.
\newblock Cancer Chemotherapy and Pharmacology. 2013;71:1013--1019.

\bibitem{benzekry2015metronomic}
Benzekry S, Pasquier E, Barbolosi D, Lacarelle B, Barlési F, André N, Ciccolini J
\newblock Metronomic reloaded: Theoretical models bringing chemotherapy into the era of precision medicine.
\newblock Seminars in Cancer Biology. 2015;35:53--61.

\bibitem{cominelli2015egfr}
Cominelli M, Grisanti S, Mazzoleni S, Branca C, Buttolo L, Furlan D, Liserre B, Bonetti MF, Medicina D, Pellegrini V, et al.
\newblock EGFR amplified and overexpressing glioblastomas and association with better response to adjuvant metronomic temozolomide.
\newblock Journal of the National Cancer Institute. 2015;107(5):djv041.
\bibitem{mercieca2018importance}
Mercieca-Bebber R, King MT, Calvert MJ, Stockler MR, Friedlander M
\newblock The importance of patient-reported outcomes in clinical trials and strategies for future optimization.
\newblock Patient Related Outcome Measures. 2018;9:353--367.

\bibitem{alfonso2020translational}
Alfonso S, Jenner AL, Craig M
\newblock Translational approaches to treating dynamical diseases through in silico clinical trials.
\newblock Chaos: An Interdisciplinary Journal of Nonlinear Science. 2020;30(12):123128.

\bibitem{dormand2018numerical}
Dormand JR
\newblock Numerical methods for differential equations: a computational approach.
\newblock Florida: CRC Press; 2018.

\bibitem{gevertz2024assessing}
Gevertz JL, Wares JR
\newblock Assessing the role of patient generation techniques in virtual clinical trial outcomes.
\newblock Bulletin of Mathematical Biology. 2024;86(10):119.

\bibitem{brown2022derisking}
Brown LV, Wagg J, Darley R, van Hateren A, Elliott T, Gaffney EA, Coles MC
\newblock De-risking clinical trial failure through mechanistic simulation.
\newblock Immunotherapy Advances. 2022;2(1):ltac017.

\bibitem{khajanchi2021impact}
Khajanchi S
\newblock The impact of immunotherapy on a glioma immune interaction model.
\newblock Chaos, Solitons \& Fractals. 2021;152:111346.

\bibitem{sturrock2015mathematical}
Sturrock M, Hao W, Schwartzbaum J, Rempala GA
\newblock A mathematical model of pre-diagnostic glioma growth.
\newblock Journal of Theoretical Biology. 2015;380:299--308.

\bibitem{bogdanska2017mathematical}
Bogda{\'n}ska MU, Bodnar M, Belmonte-Beitia J, Murek M, Schucht P, Beck J, P{\'e}rez-Garc{\'\i}a VM
\newblock A mathematical model of low grade gliomas treated with temozolomide and its therapeutical implications.
\newblock Mathematical Biosciences. 2017;288:1--13.
\bibitem{swanson2011quantifying}
Swanson KR, Rockne RC, Claridge J, Chaplain MA, Alvord Jr EC, Anderson ARA
\newblock Quantifying the role of angiogenesis in malignant progression of gliomas: in silico modeling integrates imaging and histology.
\newblock Cancer Research. 2011;71(24):7366--7375.

\bibitem{gerlee2016travelling}
Gerlee P, Nelander S
\newblock Travelling wave analysis of a mathematical model of glioblastoma growth.
\newblock Mathematical Biosciences. 2016;276:75--81.

\bibitem{stein2018mathematical}
Stein S, Zhao R, Haeno H, Vivanco I, Michor F
\newblock Mathematical modeling identifies optimum lapatinib dosing schedules for the treatment of glioblastoma patients.
\newblock PLoS Computational Biology. 2018;14(1):e1005924.

\bibitem{corwin2013toward}
Corwin D, Holdsworth C, Rockne RC, Trister AD, Mrugala MM, Rockhill JK, Stewart RD, Phillips M, Swanson KR
\newblock Toward patient-specific, biologically optimized radiation therapy plans for the treatment of glioblastoma.
\newblock PLoS One. 2013;8(11):e79115.

\bibitem{swan2018patient}
Swan A, Hillen T, Bowman JC, Murtha AD
\newblock A patient-specific anisotropic diffusion model for brain tumour spread.
\newblock Bulletin of Mathematical Biology. 2018;80:1259--1291.

\bibitem{neal2013discriminating}
Neal ML, Trister AD, Cloke T, Sodt R, Ahn S, Baldock AL, Bridge CA, Lai A, Cloughesy TF, Mrugala MM, et al.
\newblock Discriminating survival outcomes in patients with glioblastoma using a simulation-based, patient-specific response metric.
\newblock PLoS One. 2013;8(1):e51951.

\bibitem{jackson2015patient}
Jackson PR, Juliano J, Hawkins-Daarud A, Rockne RC, Swanson KR
\newblock Patient-specific mathematical neuro-oncology: using a simple proliferation and invasion tumor model to inform clinical practice.
\newblock Bulletin of Mathematical Biology. 2015;77(5):846--856.

\bibitem{gerlee2012impact}
Gerlee P, Nelander S
\newblock The impact of phenotypic switching on glioblastoma growth and invasion.
\newblock PLoS Computational Biology. 2012;8(6):e1002556.
\bibitem{aubert2006cellular}
Aubert M, Badoual M, Fereol S, Christov C, Grammaticos B
\newblock A cellular automaton model for the migration of glioma cells.
\newblock Physical Biology. 2006;3(2):93.

\bibitem{hatzikirou2012go}
Hatzikirou H, Basanta D, Simon M, Schaller K, Deutsch A
\newblock ‘Go or grow’: the key to the emergence of invasion in tumour progression?
\newblock Mathematical Medicine and Biology: A Journal of the IMA. 2012;29(1):49--65.

\bibitem{tektonidis2011identification}
Tektonidis M, Hatzikirou H, Chauvi{\`e}re A, Simon M, Schaller K, Deutsch A
\newblock Identification of intrinsic in vitro cellular mechanisms for glioma invasion.
\newblock Journal of Theoretical Biology. 2011;287:131--147.

\bibitem{saucedo2024simple}
Saucedo-Mora L, Sanz M{\'A}, Mont{\'a}ns FJ, Ben{\'\i}tez JM
\newblock A simple agent-based hybrid model to simulate the biophysics of glioblastoma multiforme cells and the concomitant evolution of the oxygen field.
\newblock Computer Methods and Programs in Biomedicine. 2024;246:108046.

\bibitem{jimenez2021mesoscopic}
Jim{\'e}nez-S{\'a}nchez J, Mart{\'\i}nez-Rubio {\'A}, Popov A, P{\'e}rez-Beteta J, Azimzade Y, Molina-Garc{\'\i}a D, Belmonte-Beitia J, Fern{\'a}ndez-Calvo G, P{\'e}rez-Garc{\'\i}a VM
\newblock A mesoscopic simulator to uncover heterogeneity and evolutionary dynamics in tumors.
\newblock PLoS Computational Biology. 2021;17(2):1--26.

\bibitem{ayala2021optimal}
Ayala-Hernández LE, Gallegos A, Schucht P, Murek M, Pérez-Romasanta L, Belmonte-Beitia J, Pérez-García VM
\newblock Optimal combinations of chemotherapy and radiotherapy in low-grade gliomas: a mathematical approach.
\newblock Journal of Personalized Medicine. 2021;11(10):1036.

\bibitem{perez2019computational}
P{\'e}rez-Garc{\'\i}a VM, Ayala-Hern{\'a}ndez LE, Belmonte-Beitia J, Schucht P, Murek M, Raabe A, Sep{\'u}lveda J
\newblock Computational design of improved standardized chemotherapy protocols for grade {II} oligodendrogliomas.
\newblock PLoS Computational Biology. 2019;15(7):e1006778.

\bibitem{delobel2023overcoming}
Delobel T, Ayala-Hernández LE, Bosque JJ, Pérez-Beteta J, Chulián S, García-Ferrer M, Piñero P, Schucht P, Murek M, Pérez-García VM
\newblock Overcoming chemotherapy resistance in low-grade gliomas: A computational approach.
\newblock PLoS Computational Biology. 2023;19(11):e1011208.

\bibitem{sujitha2023fractional}
Sujitha S, Jayakumar T, Maheskumar D
\newblock Fractional model of brain tumor with chemo-radiotherapy treatment.
\newblock Journal of Applied Mathematics and Computing. 2023;69(5):3793--3818.

\bibitem{leder2014mathematical}
Leder K, Pitter K, LaPlant Q, Hambardzumyan D, Ross BD, Chan TA, Holland EC, Michor F
\newblock Mathematical modeling of PDGF-driven glioblastoma reveals optimized radiation dosing schedules.
\newblock Cell. 2014;156(3):603--616.

\bibitem{accelrepopafterRT1}
Fogarty CE, Bergmann A
\newblock Killers creating new life: caspases drive apoptosis-induced proliferation in tissue repair and disease.
\newblock Cell Death \& Differentiation. 2017;24(8):1390--1400.

\bibitem{accelrepopafterRT2}
Zimmerman MA, Huang Q, Li F, Liu X, Li C-Y
\newblock Cell death--stimulated cell proliferation: A tissue regeneration mechanism usurped by tumors during radiotherapy.
\newblock Seminars in Radiation Oncology. 2013;23(4):288--295.
\bibitem{accelrepopafterRT3}
He S, Cheng J, Sun L, Wang Y, Wang C, Liu X, Zhang Z, Zhao M, Luo Y, Tian L, et al.
\newblock HMGB1 released by irradiated tumor cells promotes living tumor cell proliferation via paracrine effect.
\newblock Cell Death \& Disease. 2018;9(6):648.

\bibitem{vallette2019dormant}
Vallette FM, Olivier C, Lézot F, Oliver L, Cochonneau D, Lalier L, Cartron P-F, Heymann D
\newblock Dormant, quiescent, tolerant and persister cells: Four synonyms for the same target in cancer.
\newblock Biochemical Pharmacology. 2019;162:169--176.

\bibitem{rabe2020identification}
Rabé M, Dumont S, Álvarez-Arenas A, Janati H, Belmonte-Beitia J, Fernández-Calvo G, Thibault-Carpentier C, Séry Q, Chauvin C, Joalland N, et al.
\newblock Identification of a transient state during the acquisition of temozolomide resistance in glioblastoma.
\newblock Cell Death \& Disease. 2020;11(1):19.

\bibitem{dormantstateandback}
Yeh AC, Ramaswamy S
\newblock Mechanisms of cancer cell dormancy—another hallmark of cancer?
\newblock Cancer Research. 2015;75(23):5014--5022.

\bibitem{dahlrot2021prognostic}
Dahlrot RH, Bangs{\o} JA, Petersen JK, Rosager AM, S{\o}rensen MD, Reifenberger G, Hansen S, Kristensen BW
\newblock Prognostic role of Ki-67 in glioblastomas excluding contribution from non-neoplastic cells.
\newblock Scientific Reports. 2021;11(1):17918.

\bibitem{lomax2013biological}
Lomax ME, Folkes LK, O'Neill P
\newblock Biological consequences of radiation-induced DNA damage: relevance to radiotherapy.
\newblock Clinical Oncology. 2013;25(10):578--585.

\bibitem{bacteriapersisters}
Michiels JE, Van den Bergh B, Verstraeten N, Michiels J
\newblock Molecular mechanisms and clinical implications of bacterial persistence.
\newblock Drug Resistance Updates. 2016;29:76--89.

\bibitem{ramirez2016diverse}
Ramirez M, Rajaram S, Steininger RJ, Osipchuk D, Roth MA, Morinishi LS, Evans L, Ji W, Hsu C-H, Thurley K, et al.
\newblock Diverse drug-resistance mechanisms can emerge from drug-tolerant cancer persister cells.
\newblock Nature Communications. 2016;7(1):10690.
\bibitem{paramgeneracionpersisters}
Sharma SV, Lee DY, Li B, Quinlan MP, Takahashi F, Maheswaran S, McDermott U, Azizian N, Zou L, Fischbach MA, et al.
\newblock A chromatin-mediated reversible drug-tolerant state in cancer cell subpopulations.
\newblock Cell. 2010;141(1):69--80.

\bibitem{oren2021cycling}
Oren Y, Tsabar M, Cuoco MS, Amir-Zilberstein L, Cabanos HF, Hütter J-C, Hu B, Thakore PI, Tabaka M, Fulco CP, et al.
\newblock Cycling cancer persister cells arise from lineages with distinct programs.
\newblock Nature. 2021;596(7873):576--582.

\bibitem{rambow2018toward}
Rambow F, Rogiers A, Marin-Bejar O, Aibar S, Femel J, Dewaele M, Karras P, Brown D, Chang YH, Debiec-Rychter M, et al.
\newblock Toward minimal residual disease-directed therapy in melanoma.
\newblock Cell. 2018;174(4):843--855.

\bibitem{persisters1}
Roesch A, Fukunaga-Kalabis M, Schmidt EC, Zabierowski SE, Brafford PA, Vultur A, Basu D, Gimotty P, Vogt T, Herlyn M
\newblock A temporarily distinct subpopulation of slow-cycling melanoma cells is required for continuous tumor growth.
\newblock Cell. 2010;141(4):583--594.

\bibitem{hangauer2017drug}
Hangauer MJ, Viswanathan VS, Ryan MJ, Bole D, Eaton JK, Matov A, Galeas J, Dhruv HD, Berens ME, Schreiber SL, et al.
\newblock Drug-tolerant persister cancer cells are vulnerable to GPX4 inhibition.
\newblock Nature. 2017;551(7679):247--250.
\bibitem{boyd2018identification}
Boyd AL, Aslostovar L, Reid J, Ye W, Tanasijevic B, Porras DP, Shapovalova Z, Almakadi M, Foley R, Leber B, et al.
\newblock Identification of chemotherapy-induced leukemic-regenerating cells reveals a transient vulnerability of human AML recurrence.
\newblock Cancer Cell. 2018;34(3):483--498.

\bibitem{ho2016evolution}
Ho T-C, LaMere M, Stevens BM, Ashton JM, Myers JR, O’Dwyer KM, Liesveld JL, Mendler JH, Guzman M, Morrissette JD, et al.
\newblock Evolution of acute myelogenous leukemia stem cell properties after treatment and progression.
\newblock Blood. 2016;128(13):1671--1678.

\bibitem{segura2021tumor}
Segura-Collar B, Garranzo-Asensio M, Herranz B, Hernández-SanMiguel E, Cejalvo T, Casas BS, Matheu A, Pérez-Núñez Á, Sepúlveda-Sánchez JM, Hernández-Laín A, et al.
\newblock Tumor-derived pericytes driven by EGFR mutations govern the vascular and immune microenvironment of gliomas.
\newblock Cancer Research. 2021;81(8):2142--2156.

\bibitem{gargini2020idh}
Gargini R, Segura-Collar B, Herranz B, García-Escudero V, Romero-Bravo A, Núñez FJ, García-Pérez D, Gutiérrez-Guamán J, Ayuso-Sacido A, Seoane J, et al.
\newblock The IDH-TAU-EGFR triad defines the neovascular landscape of diffuse gliomas.
\newblock Science Translational Medicine. 2020;12(527):eaax1501.

\bibitem{chemodamage}
Mills CC, Kolb EA, Sampson VB.
\newblock Development of chemotherapy with cell-cycle inhibitors for adult and pediatric cancer therapy.
\newblock Cancer Research. 2018;78(2):320--325.

\bibitem{radiodamage}
Pawlik TM, Keyomarsi K.
\newblock Role of cell cycle in mediating sensitivity to radiotherapy.
\newblock Int J Radiat Oncol Biol Phys. 2004;59(4):928--942.

\bibitem{hirose2005akt}
Hirose Y, Katayama M, Mirzoeva OK, Berger MS, Pieper RO.
\newblock Akt activation suppresses Chk2-mediated, methylating agent-induced G2 arrest and protects from temozolomide-induced mitotic catastrophe and cellular senescence.
\newblock Cancer Research. 2005;65(11):4861--4869.

\bibitem{vakifahmetoglu2008death}
Vakifahmetoglu H, Olsson M, Zhivotovsky B.
\newblock Death through a tragedy: mitotic catastrophe.
\newblock Cell Death Differ. 2008;15(7):1153--1162.

\bibitem{programmedcelldeathglioma}
Jakubowicz-Gil J, Bkadziul D, Langner E, Wertel I, Zajkac A, Rzeski W.
\newblock Temozolomide and sorafenib as programmed cell death inducers of human glioma cells.
\newblock Pharmacol Rep. 2017;69(4):779--787.

\bibitem{joiner2025basic}
Joiner MC, van der Kogel A.
\newblock Basic Clinical Radiobiology.
\newblock Florida: CRC Press; 2009.
\bibitem{adamski2017dormant}
Adamski V, Hempelmann A, Flüh C, Lucius R, Synowitz M, Hattermann K, Held-Feindt J.
\newblock Dormant glioblastoma cells acquire stem cell characteristics and are differentially affected by temozolomide and AT101 treatment.
\newblock Oncotarget. 2017;8(64):108064.


\bibitem{otero2022dynamics}
García Otero J, Álvarez-Arenas Alcamí A, Belmonte-Beitia J  
\newblock Dynamics and analysis of a mathematical model of neuroblastoma treated with Celyvir.
\newblock Applied Mathematical Modelling. 2022;110:131--148.

\bibitem{otero2024dynamics}
García Otero J, Bodzioch M, Belmonte-Beitia J  
\newblock On the dynamics and optimal control of a mathematical model of neuroblastoma and its treatment: Insights from a mathematical model.
\newblock Mathematical Models and Methods in Applied Sciences. 2024;34(07):1235--1278.


\bibitem{neuropharmacokinetics}
Portnow J, Badie B, Chen M, Liu A, Blanchard S, Synold TW.
\newblock The neuropharmacokinetics of temozolomide in patients with resectable brain tumors: potential implications for the current approach to chemoradiation.
\newblock Clin Cancer Res. 2009;15(22):7092--7098.

\bibitem{ballesta2014multiscale}
Ballesta A, Zhou Q, Zhang X, Lv H, Gallo JM.
\newblock Multiscale design of cell-type-specific pharmacokinetic/pharmacodynamic models for personalized medicine: application to temozolomide in brain tumors.
\newblock CPT Pharmacometrics Syst Pharmacol. 2014;3(4):1--11.

\bibitem{akkari2020dynamic}
Akkari L, Bowman RL, Tessier J, Klemm F, Handgraaf SM, de Groot M, Quail DF, Tillard L, Gadiot J, Huse JT, et al.
\newblock Dynamic changes in glioma macrophage populations after radiotherapy reveal CSF-1R inhibition as a strategy to overcome resistance.
\newblock Sci Transl Med. 2020;12(552):eaaw7843.

\bibitem{rutter2017mathematical}
Rutter EM, Stepien TL, Anderies BJ, Plasencia JD, Woolf EC, Scheck AC, Turner GH, Liu Q, Frakes D, Kodibagkar V, et al.
\newblock Mathematical analysis of glioma growth in a murine model.
\newblock Sci Rep. 2017;7(1):2508.

\bibitem{yuan2018abt}
Yuan AL, Ricks CB, Bohm AK, Lun X, Maxwell L, Safdar S, Bukhari S, Gerber A, Sayeed W, Bering EA, et al.
\newblock ABT-888 restores sensitivity in temozolomide-resistant glioma cells and xenografts.
\newblock PLoS One. 2018;13(8):e0202860.

\bibitem{yang2021enhanced}
Yang TC, Liu SJ, Lo WL, Chen SM, Tang YL, Tseng YY.
\newblock Enhanced anti-tumor activity in mice with temozolomide-resistant human glioblastoma cell line-derived xenograft using SN-38-incorporated polymeric microparticle.
\newblock Int J Mol Sci. 2021;22(11):5557.

\bibitem{gupta2014discordant}
Gupta SK, Mladek AC, Carlson BL, Boakye-Agyeman F, Bakken KK, Kizilbash SH, Schroeder MA, Reid J, Sarkaria JN.
\newblock Discordant in vitro and in vivo chemopotentiating effects of the PARP inhibitor veliparib in temozolomide-sensitive versus-resistant glioblastoma multiforme xenografts.
\newblock Clin Cancer Res. 2014;20(14):3730--3741.

\bibitem{perez2015delay}
Pérez-García VM, Bogdanska M, Martínez-González A, Belmonte-Beitia J, Schucht P, Pérez-Romasanta LA.
\newblock Delay effects in the response of low-grade gliomas to radiotherapy: a mathematical model and its therapeutical implications.
\newblock Math Med Biol. 2015;32(3):307--329.

\bibitem{cho2020pharmacokinetic}
Cho HY, Swenson S, Thein TZ, Wang W, Wijeratne NR, Marín-Ramos NI, Katz JE, Hofman FM, Schönthal AH, Chen TC.
\newblock Pharmacokinetic properties of the temozolomide perillyl alcohol conjugate (NEO212) in mice.
\newblock Neurooncol Adv. 2020;2(1):vdaa160.

\bibitem{wu2022integrating}
Wu C, Lorenzo G, Hormuth DA, Lima EABF, Slavkova KP, DiCarlo JC, Virostko J, Phillips CM, Patt D, Chung C, et al.
\newblock Integrating mechanism-based modeling with biomedical imaging to build practical digital twins for clinical oncology.
\newblock Biophys Rev. 2022;3(2):021304.

\bibitem{oraiopoulou2017vitro}
Oraiopoulou ME, Tzamali E, Tzedakis G, Vakis A, Papamatheakis J, Sakkalis V.
\newblock In vitro/in silico study on the role of doubling time heterogeneity among primary glioblastoma cell lines.
\newblock Biomed Res Int. 2017;2017:8569328.

\bibitem{IVIS}
Zheng J, Xu L, Zhou H, Zhang W, Chen Z.
\newblock Quantitative analysis of cell tracing by in vivo imaging system.
\newblock J Huazhong Univ Sci Technol Med Sci. 2010;30(4):541--545.
\bibitem{creed2020matsurv}
Creed JH, Gerke TA, Berglund AE
\newblock MatSurv: Survival analysis and visualization in {MATLAB}.
\newblock Journal of Open Source Software. 2020;5(46):1830.

\bibitem{Ode2}
León-Triana OT, Pérez-Martínez AA, Ramírez-Orellana M, Pérez-García VM
\newblock Dual-target {CAR}-{T}s with on-and off-tumour activity may override immune suppression in solid cancers: A mathematical proof of concept.
\newblock Cancers. 2021;13(4):703.

\bibitem{jackson2019mid}
Jackson WC, Suresh K, Maurino C, Feng M, Cuneo KC, Ten Haken RK, Lawrence TS, Schipper MJ, Owen D
\newblock A mid-treatment break and reassessment maintains tumor control and reduces toxicity in patients with hepatocellular carcinoma treated with stereotactic body radiation therapy.
\newblock Radiotherapy and Oncology. 2019;141:101--107.

\bibitem{sun2018ki}
Sun X, Kaufman PD
\newblock Ki-67: more than a proliferation marker.
\newblock Chromosoma. 2018;127:175--186.

\bibitem{victoria2015incorporating}
Victoria YY, Nguyen D, Pajonk F, Kupelian P, Kaprealian T, Selch M, Low DA, Sheng K
\newblock Incorporating cancer stem cells in radiation therapy treatment response modeling and the implication in glioblastoma multiforme treatment resistance.
\newblock International Journal of Radiation Oncology Biology Physics. 2015;91(4):866--875.

\bibitem{panetta2003mechanistic}
Panetta JC, Kirstein MN, Gajjar AJ, Nair G, Fouladi M, Stewart CF
\newblock A mechanistic mathematical model of temozolomide myelosuppression in children with high-grade gliomas.
\newblock Mathematical biosciences. 2003;186(1):29--41.

\bibitem{perez2024modelling}
Pérez-Aliacar M, Ayensa-Jiménez J, Randjelović T, Ochoa I, Doblaré M
\newblock Modelling glioblastoma resistance to temozolomide. A mathematical model to simulate cellular adaptation in vitro.
\newblock Computers in Biology and Medicine. 2024;180:108866.

\bibitem{yan20173d}
Yan H, Romero-López M, Benitez LI, Di K, Frieboes HB, Hughes CW, Bota DA, Lowengrub JS
\newblock 3D mathematical modeling of glioblastoma suggests that transdifferentiated vascular endothelial cells mediate resistance to current standard-of-care therapy.
\newblock Cancer research. 2017;77(15):4171--4184.
\bibitem{tafech2023characterization}
Tafech A, Jacquet P, Beaujean C, Fertin A, Usson Y, Stéphane A
\newblock Characterization of the intracellular acidity regulation of brain tumor cells and consequences for therapeutic optimization of temozolomide.
\newblock Biology. 2023;12(9):1221.

\bibitem{stephanou2019ph}
Stéphane A, Ballesta A
\newblock pH as a potential therapeutic target to improve temozolomide antitumor efficacy: A mechanistic modeling study.
\newblock Pharmacology Research \& Perspectives. 2019;7(1):e00454.

\bibitem{oraiopoulou2024temozolomide}
Oraiopoulou M-E, Tzamali E, Psycharakis SE, Tzedakis G, Makatounakis T, Manolitsi K, Drakos E, Vakis AF, Zacharakis G, Papamatheakis J, others
\newblock The Temozolomide--Doxorubicin paradox in Glioblastoma in vitro--in silico preclinical drug-screening.
\newblock Scientific Reports. 2024;14(1):3759.

\bibitem{surendran2023agent}
Surendran A, Jenner AL, Karimi E, Fiset B, Quail DF, Walsh LA, Craig M
\newblock Agent-based modelling reveals the role of the tumor microenvironment on the short-term success of combination temozolomide/immune checkpoint blockade to treat glioblastoma.
\newblock The Journal of Pharmacology and Experimental Therapeutics. 2023;387(1):66--77.

\bibitem{houy2018optimal}
Houy N, Le Grand F
\newblock Optimal dynamic regimens with artificial intelligence: The case of temozolomide.
\newblock PLoS One. 2018;13(6):e0199076.

\bibitem{zade2020reinforcement}
Zade AE, Haghighi SS, Soltani M
\newblock Reinforcement learning for optimal scheduling of Glioblastoma treatment with Temozolomide.
\newblock Computer methods and programs in biomedicine. 2020;193:105443.

\bibitem{sorribes2020mitigating}
Sorribes IC, Handelman SK, Jain HV
\newblock Mitigating temozolomide resistance in glioblastoma via DNA damage-repair inhibition.
\newblock Journal of the Royal Society Interface. 2020;17(162):20190722.

\bibitem{falco2021silico}
Falco J, Agosti A, Vetrano IG, Bizzi A, Restelli F, Broggi M, Schiariti M, DiMeco F, Ferroli P, Ciarletta P, others
\newblock In silico mathematical modelling for glioblastoma: a critical review and a patient-specific case.
\newblock Journal of clinical medicine. 2021;10(10):2169.

\bibitem{brady2019mathematical}
Brady R, Enderling H
\newblock Mathematical models of cancer: when to predict novel therapies, and when not to.
\newblock Bulletin of mathematical biology. 2019;81:3722--3731.

\bibitem{italia2022optimal}
Italia M, Dercole F, Lucchetti R
\newblock Optimal chemotherapy counteracts cancer adaptive resistance in a cell-based, spatially-extended, evolutionary model.
\newblock Physical Biology. 2022;19(2):026004.
\bibitem{wang2024virtual}
Wang H, Arulraj T, Ippolito A, Popel AS
\newblock From virtual patients to digital twins in immuno-oncology: lessons learned from mechanistic quantitative systems pharmacology modeling.
\newblock NPJ Digital Medicine. 2024;7(1):189.

\bibitem{van2010absence}
Van Nifterik KA, Van Den Berg J, Van Der Meide WF, Ameziane N, Wedekind LE, Steenbergen RDM, Leenstra S, Lafleur MVM, Slotman BJ, Stalpers LJA, others
\newblock Absence of the MGMT protein as well as methylation of the MGMT promoter predict the sensitivity for temozolomide.
\newblock British journal of cancer. 2010;103(1):29--35.

\bibitem{martinez2012hypoxic}
Martínez-González A, Calvo GF, Pérez Romasanta LA, Pérez-García VM
\newblock Hypoxic cell waves around necrotic cores in glioblastoma: a biomathematical model and its therapeutic implications.
\newblock Bulletin of mathematical biology. 2012;74:2875--2896.

\bibitem{celora2024characterising}
Celora GL, Nixson R, Pitt-Francis JM, Maini PK, Byrne HM
\newblock Characterising cancer cell responses to cyclic hypoxia using mathematical modelling.
\newblock Bulletin of Mathematical Biology. 2024;86(12):145.

\bibitem{spinicci2025modelling}
Spinicci K, Powathil G, Stéphane A
\newblock Modelling the impact of HIF on metabolism and the extracellular matrix: consequences for tumour growth and invasion.
\newblock Bulletin of Mathematical Biology. 2025;87(2):1--26.



\end{thebibliography}
\end{document}